\documentclass[useAMS,usenatbib,usegraphicx]{mn2e}
\usepackage{graphicx,times,epsfig}

\newcommand{\kmsend}{\mbox{km s$^{-1}$}}
 \newcommand{\kms}{\mbox{km s$^{-1}$ }}

\newcommand{\msun}{\mbox{M$_{\sun}$ }}
\newcommand{\msunend}{\mbox{M$_{\sun}$}}

\newcommand{\msunyr}{\mbox{M$_{\sun}$yr$^{-1}$ }}
\newcommand{\msunyrend}{\mbox{M$_{\sun}$yr$^{-1}$}}
\newcommand{\htwo}{\mbox{H$_2$}}

\newcommand{\z}{\mbox{$z$}}

\newcommand{\zsim}{\mbox{$z\sim$ }}

\newcommand{\mdyn}{\mbox{M$_{\rm dyn}$}}

 \newcommand{\sef}{\mbox{$S_{\rm 850}$}}

\newcommand{\sunrise}{\mbox{\sc sunrise}}
\newcommand{\gadget}{\mbox{\sc gadget-3}}
\newcommand{\gadgettwo}{\mbox{\sc gadget-2}}

\newcommand{\mappings}{\mbox{\sc mappingsiii}}
\newcommand{\turtlebeach}{\mbox{\sc turtlebeach}}

\title[Molecular Gas in SMGs]{The
 Star-Forming Molecular Gas in High Redshift Submillimeter Galaxies}

\author[Narayanan, Cox, Hayward, Younger \& Hernquist]{Desika\,
  Narayanan\thanks{E-mail: dnarayanan@cfa.harvard.edu}\thanks{CfA
    Fellow}, Thomas\, J.\, Cox\thanks{W.M. Keck Postdoctoral Fellow},
  Christopher\, C.\, Hayward, Joshua\, D.\, Younger, \and and Lars\,
  Hernquist\\Harvard-Smithsonian Center for Astrophysics, 60 Garden
  St., Cambridge, Ma 02138}
\begin{document}

\date{Submitted to MNRAS}

\pagerange{\pageref{firstpage}--\pageref{lastpage}} \pubyear{2009}

\maketitle

\label{firstpage}

\begin{abstract}

We present a model for the CO molecular line emission from high
redshift Submillimeter Galaxies (SMGs). By combining hydrodynamic
simulations of gas rich galaxy mergers with the polychromatic
radiative transfer code, \sunrise, and the 3D non-LTE molecular
line radiative transfer code, \turtlebeach, we show that if SMGs
are typically a transient phase of major mergers, their observed
compact CO spatial extents, broad line widths, and high excitation
conditions (CO SED) are naturally explained. In this sense, SMGs can
be understood as scaled-up analogs to local ULIRGs. We utilize these
models to investigate the usage of CO as an indicator of physical
conditions. We find that care must be taken when applying standard
techniques. The usage of CO line widths as a dynamical mass estimator
from SMGs can possibly overestimate the true enclosed mass by a factor
$\sim$1.5-2. At the same time, assumptions of line ratios of unity
from CO J=3-2 (and higher lying lines) to CO (J=1-0) will oftentimes
lead to underestimates of the inferred gas mass.  We provide tests for
these models by outlining predictions for experiments which are
imminently feasible with the current generation of bolometer arrays
and radio-wave spectrometers.

\end{abstract}


\section{Introduction}

Understanding the origin and evolution of active galaxy populations at
cosmological redshifts remains an outstanding problem.  Of particular
interest are a population of submillimeter-luminous galaxies
discovered via deep, blind surveys with SCUBA on the JCMT at a median
redshift of \zsim 2 \citep{bar98,hug98,cha03a}.  These Submillimeter
Galaxies (SMGs) are empirically defined with the flux limit S$_{850
  \mu m} \ga$ 5-6 mJy\footnote{This empirical definition is for
  blank-field surveys at current sensitivity limits.}
  \citep[e.g. ][]{bla02}, and appear to be a highly clustered
  population of galaxies forming stars at prodigious rates \citep[SFR
    $\ga$ 10$^3$ \msunyr;
  ][]{bla02,bla04,swi04,men07,kov06,val07,cop08b}. In addition to
  undergoing prodigious star formation, some SMGs additionally host
  heavily obscured AGN
  \citep[e.g.][]{ivi02,ale05a,ale05b,ale08,bor05}. These galaxies may
  contribute a substantial fraction of the cosmic star formation
  density \citep{bla99,bla02}, as well as serve as prime candidates
  for studying the coevolution of black hole growth and star formation
  at an epoch of heightened galaxy formation and evolution.

Many open questions remain regarding the physical nature of SMGs. Are
they isolated galaxies, or galaxy mergers? What is their potential
place in an evolutionary sequence? Are they high-redshift analogs to
local infrared luminous galaxies (e.g. ULIRGs)? What are their typical
gaseous and stellar masses?

Emission from the star forming molecular interstellar medium (ISM) has
the potential to elucidate some of these questions. For example,
$^{12}$CO (J=1-0; hereafter, CO) emission can serve as a measure both
for the total amount of \htwo \ molecular gas available for future
star formation \citep[e.g.][]{dow98,sol91,gre05}, as well as previous
stellar mass assembly by serving as a dynamical mass tracer
\citep[e.g.][]{tac06,bou07,ho07,nar08b}. The high resolution images
available from millimeter-wave interferometry can reveal source sizes
and morphologies
\citep[e.g.][]{gen03,tac06,tac08,you08b,ion09}. Additionally, the
rotational ladder from \htwo \ tracers such as CO reveal the thermal
conditions and mean densities of the star forming ISM
\citep{wei05,wei07}.  CO has been detected routinely in high redshift
SMGs for over a decade \citep[e.g. ][and references
  therein]{fra98,fra99,ner03,gen03,gre05,tac06,tac08}.  During this
time, a number of pioneering papers have afforded the community a
wealth of CO data from \zsim 2 SMGs. The general molecular emission
properties from SMGs can be summarized as follows.

{\it \htwo \ Gas Masses:} SMGs are incredibly gas rich, with massive
\htwo \ gas reservoirs. Extrapolation from typical CO (J=3-2) rest
frame measurements from SMGs suggests \htwo \ masses of order
10$^{10}$-10$^{11}$ \msunend, making these some of the most molecular
gas rich galaxies in the Universe
\citep{gre05,sol05,tac06,car06,tac08,cop08}. An analysis of
semi-analytic models of SMG formation by \citet{swi08} found good
agreement between the cold gas mass in their simulations and the
observed values.

{\it Line Widths:} The CO line widths in SMGs are typically broad with
median line width ranging from $\sim$600-800 \kms
\citep[FWHM;][]{gre05,car06,tac06,cop08,tac08,ion09}. Comparisons to
CO-detected quasars at comparable redshifts have had conflicting
results.  Analysis of literature data by \citet{car06} has suggested
that SMGs typically have broader line widths than \zsim 2 quasars by a
factor of $\sim$2.5 with KS tests indicating the two are not drawn
from the same parent population at the $>$99\% confidence level.  On
the other hand, new detections by \citet{cop08} have found rather
similar CO line widths from quasars and SMGs, and KS tests showing
that the two arise from the same parent population at the $>$95\%
confidence level.


{\it Images:} Interferometric CO imaging of SMGs has shown the spatial
distribution of most SMGs to be relatively compact \citep[R$_{\rm
    hwhm}$ $\la$ 1-2 kpc;][]{tac06}. Some individual sources have
shown rather extended emission \citep[e.g.][]{gen03}. Curiously,
images have shown evidence for compact disk-like motion in at least a
few sources \citep{gen03,tac06}, as well as extended emission in what
appears to be interacting/merging galaxies
\citep[e.g.][]{ivi01,tac06}.

{\it Excitation Properties:} While very few SMGs have been observed in
multiple CO lines, the existing data suggests that these galaxies
exhibit relatively high molecular excitation conditions. The CO line
spectral energy distributions (CO SEDs; alternatively known as CO
rotational ladders) are seen to typically turn over at the J=5 level
\citep[$\sim$83 K above ground; ][]{wei05,wei07,gre05,hai06}.  That
said, few objects have been detected in their rest-frame CO (J=1-0)
line \citep{hai06}. As \htwo \ masses are typically inferred from CO
(J=1-0) measurements, the excitation properties of higher lying
levels (e.g. line ratios) with respect to the ground state transition
are crucial to constrain for the purposes of deriving molecular gas
masses.

While the information provided by CO observations of \zsim 2 SMGs is
indeed invaluable, questions remain regarding the interpretation of
many of the various aforementioned observational characteristics of
these sources. For example, what is the true relationship between CO
line widths from SMGs and quasars?  Is there an evolution in the CO
line widths of SMGs? How reliably can the CO line width be used as
dynamical mass tracer in these galaxies?  How can observations of
higher lying CO lines (in the rest frame) be extrapolated to the rest
frame CO (J=1-0) luminosity in order to derive an \htwo \ gas mass
\citep[e.g.][]{hai06}? Can the observed CO emission line widths,
excitation properties, and images be explained by a merger-driven
scenario for SMG formation and evolution \citep[e.g. ``scaled up
  ULIRGs'';][]{tac06,tac08}?  It is clear that a theoretical
interpretation behind the molecular line emission properties may be
valuable for elucidating some of the aforementioned issues, as well as
providing interpretation for forthcoming observations.

Along with providing interpretation for observations, theoretical
calculations of CO emission from simulated SMGs can provide direct
tests of models of SMG formation and evolution.  In \citet{nar09}, we
presented a merger-driven model for the formation of SMGs which
reproduced the full range of observed 850 \micron \ fluxes from SMGs,
the optical-mm wave SED, and characteristic stellar, black hole, and
dark matter masses. Comparing the simulated molecular gas properties
of these model SMGs to the extensive data sets in the literature
provides a strict test of the models. In this paper, we investigate
the CO emission properties from SMGs by combining these SMG formation
models with 3D non-local thermodynamic equilibrium (LTE) molecular
line radiative transfer calculations \citep{nar06a,nar08a}. The goals
of this paper are to: (a) provide direct tests of merger-driven
formation mechanisms for SMGs by comparing the simulated CO emission
from the models of \citet{nar09} to observations, and, given a
sufficient correspondence between models and observations, (b) provide
interpretation for existing and future observational data.

This paper is organized as follows: in \S~\ref{section:methods}, we
describe our numerical methods for our hydrodynamic, molecular line
radiative transfer, and polychromatic SED radiative transfer
simulations. In \S~\ref{section:evolution}, we summarize the evolution
of the submillimeter, $B$-band, and \htwo \ properties of our
simulated galaxies. In \S~\ref{section:morphology}, we study the CO
morphology, molecular disk formation and CO spatial extents in
SMGs. In \S~\ref{section:linewidths}, we explore the origin of the
broad observed CO line widths. We use these results to investigate the
usage of CO as a dynamical mass tracer in \S~\ref{section:mdyn}, and
analyze the CO excitation in SMGs in \S~\ref{section:excitation}.  In
\S~\ref{section:discussion}, we provide imminently testable
observational predictions, and in \S~\ref{section:summary}, we
summarize.

\section{Numerical Methods}
\label{section:methods}

\begin{figure}
\hspace{-2cm}
\vspace{-1cm}
\includegraphics[scale=0.4,angle=90]{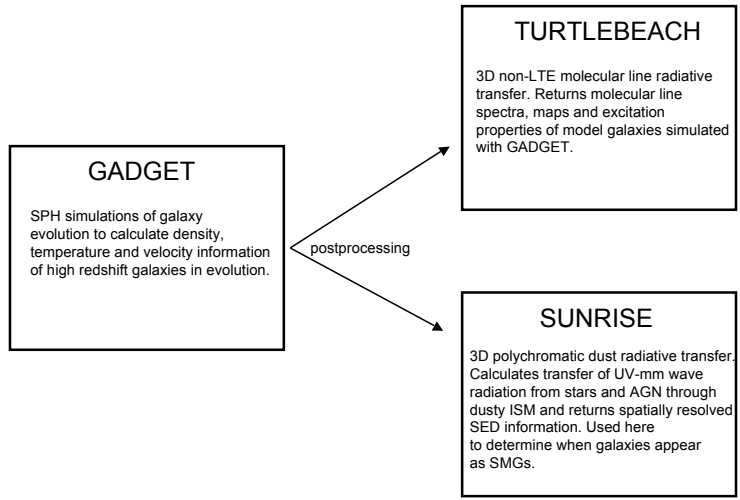}
\caption{Flowchart summarizing methodology. The galaxies are simulated
  hydrodynamically using the SPH code, \gadget \ \citep{spr05b}. The
  simulated molecular line emission and SEDs are then calculated in
  post-processing utilizing the radiative transfer codes \turtlebeach
  \ \citep{nar08a} and \sunrise \ \citep{jon09},
  respectively.\label{figure:flowchart}}

\end{figure}

Generically, our methodology involves three steps. First, we simulate
the hydrodynamic evolution of galaxies utilizing the smoothed-particle
hydrodynamics (SPH) code, \gadget \ \citep{spr05b}. We then calculate
the molecular line emission properties and SEDs from these simulated
galaxies in post-processing utilizing the radiative transfer codes
\turtlebeach \ \citep{nar08a} and \sunrise \ \citep{jon09}. This method
is summarized in the flowchart presented in
Figure~\ref{figure:flowchart}. In this section, we present the details
of the hydrodynamic and radiative transfer methods.

\subsection{Hydrodynamics}
\label{section:hydro}

\begin{table*}
\label{table:ICs}
\centering
\begin{minipage}{100mm}
\caption{SMGs are ordered with decreasing halo mass. Column 1 is the
  name of the model used in this work.  Columns 2 \& 3 are initial
  orientations for disk 1, Columns 4 \& 5 are for disk 2. Column 6
  gives the virial velocity of the progenitors, Column 7 gives their
  halo masses, and Column 8 the mass ratio of the merger. Please see
  \citet{nar09} for generalized results regarding the physical
  properties of these galaxies.}
\begin{tabular}{@{}cccccccc@{}}
\hline Model & $\theta_1$ &$\phi_1$ & $\theta_2$ & $\phi_2$&
  V$_{\rm c}$ & $M_{\rm DM}$ & Mass Ratio \\
&&&&&(\kmsend)&\msun &\\
\hline
SMG1 & 30  & 60 & -30 & 45 & 500:500 & 2.5$\times$10$^{13}$ & 1:1 \\
SMG2 &  360  & 60 & 150 & 0 & 500:500 &  2.5$\times$10$^{13}$ & 1:1 \\
SMG3 & -109  & -30 & 71 & -30 & 500:500 &  2.5$\times$10$^{13}$ &  1:1 \\
SMG4 & 30  & 60 & -30 & 45 & 500:320 & 1.6$\times$10$^{13}$ & 1:3 \\
SMG5 &  360  & 60 & 150 & 0 & 500:320 &  1.6$\times$10$^{13}$ & 1:3 \\
SMG6 & -109  & -30 & 71 & -30 & 500:320 &  1.6$\times$10$^{13}$ &  1:3 \\
SMG7 & 30  & 60 & -30 & 45 & 500:225 & 1.4$\times$10$^{13}$ & 1:12 \\
SMG8 &  360  & 60 & 150 & 0 & 500:225 &  1.4$\times$10$^{13}$ & 1:12 \\
SMG9 & -109  & -30 & 71 & -30 & 500:225 &  1.4$\times$10$^{13}$ &  1:12 \\
SMG10 & 30  & 60 & -30 & 45 & 320:320 & 6.6$\times$10$^{12}$ & 1:1 \\
SMG11 &  360  & 60 & 150 & 0 & 320:320 &  6.6$\times$10$^{12}$ & 1:1 \\
SMG12 & -109  & -30 & 71 & -30 & 320:320 &  6.6$\times$10$^{12}$ &  1:1 \\
SMG13 & 30  & 60 & -30 & 45 & 225:225 & 2.35$\times$10$^{12}$ & 1:1 \\
\hline
\end{tabular}
\end{minipage}
\end{table*}

A major goal of our program is to understand the evolution of model
SMGs while remaining constrained to observations for physical input to
the simulations. When data for SMGs is unavailable, we turn to
observational constraints from the Galaxy or local starbursts.  We
consider the formation of SMGs in gas rich binary galaxy mergers at
high redshift. This is motivated by observed radio, CO, and optical
morphologies of SMGs which appear to show signs of interactions
\citep[e.g.][]{cha03b, tac06, tac08}, as well as theoretical models
which demonstrate that mergers serve as an efficient means of
triggering nuclear starbursts
\citep{bar91,bar96,mih94a,mih96,spr05a}. Moreover, simulations by
\citet{nar09} have suggested that isolated galaxies and minor mergers
(mass ratio $<<$1:10) are unlikely to result in a \sef$>$5 mJy SMG.

The hydrodynamic simulations were performed with the smoothed-particle
hydrodynamics code, \gadget\footnote{The main improvement in \gadget
  \ over \gadgettwo \ \citep[described by ][]{spr05b} is better load
  balancing on parallel processors.} \ \citep{spr05b} which utilizes a
fully conservative SPH formalism \citep{spr02}. The hydrodynamic
simulations include prescriptions for radiative cooling of the gas
\citep{kat96,dav99}, and a multiphase ISM in which cold clouds are
considered to be in pressure equilibrium with hot gas
\citep{mck77,spr03a}. This multi-phase ISM is implemented such that
cold clouds grow through radiative cooling of hot gas, and heating
from star formation can evaporate cold clouds \citep{spr05a}.  The
effect of supernovae on the ISM is treated via an effective equation
of state (EOS). Here, we employ the full multi-phase EOS where
supernovae-driven pressure optimally maintain disk-stability
\citep[$q_{\rm EOS}$=1; For more details, see Figure 4 of ][]{spr05a}.

Star formation occurs following a volumetric generalization of the
Kennicutt-Schmidt relation, SFR $\propto$ $n^{1.5}$
\citep{ken98a,ken98b,ken07,sch59,spr03a}, which results in disks
consistent with the local surface-density scaling relations
\citep{cox06a}. The star formation timescale is chosen such that
isolated disk models are consistent with the normalization of the
local Kennicutt-Schmidt relations. While there are rather few
measurements of observed star formation rate (SFR) relations at
cosmological redshifts, tentative evidence exists that \zsim 2
galaxies may lie on the present-day Kennicutt-Schmidt relation
\citep{bou07}. Indeed, this is theoretically favorable as the relation
SFR$\sim$n$^{1.5}$ may be understood in terms of free-fall time
arguments which are redshift invariant.

Black holes are included in the simulations as sink particles which
accrete following a Bondi-Hoyle-Lyttleton parameterization according to
the local gas density and sound speed \citep{bon44,bon52}. The
bolometric luminosity of the black hole is set at $L_{\rm
  bol}$=$\epsilon \dot{M}c^2$, where $\epsilon$=0.1. Feedback from the
black hole(s) is modeled such that a fixed fraction of this luminosity
(here, 5\%) couples thermally and isotropically to the surrounding
ISM. This coupling efficiency is set to match the normalization of
the local $M-\sigma$ relation \citep{dim05,spr05a}.

 The progenitor disk galaxies are initialized with a \citet{her90}
 dark matter halo profile \citep[for details, see ][]{spr05a} with
 virial properties scaled to be appropriate for \zsim 3 \citep[such
   that they may represent galaxies undergoing a major merger by \zsim
   2;][]{bul01,rob06b}. The progenitors begin with an initial gas
 fraction of $f_g$=0.8, resulting in a gas fraction of $f_{\rm g}
 \sim$20-40\% by the time the galaxies approach final
 coalescence. This is comparable to recent measurements by
 \citet{bou07} and \citet{tac08} who find tentative evidence for gas
 fractions in \zsim 2 SMGs of $\sim$ 40\%.

Here, we consider the \citet{nar09} series of 12 models, varying total
mass, mass ratio, and orbit.  We include one additional lower mass
merger which will not make an SMG (model SMG13), though will be useful
for predictions of lower flux galaxies
(\S~\ref{section:discussion}). The initial conditions of the merger
simulations are summarized in Table~\ref{table:ICs}. The primary
progenitors are initialized with circular velocities ranging from
$V_c$=320-500 \kmsend, similar to measurements tabulated by
\citet{tac08}. This corresponds to halo masses of order $M_{\rm DM}
\sim$10$^{12}$-10$^{13}$\msunend, comparable to those inferred by
clustering measurements \citep{bla04,swi08}. We model the formation of
SMGs in 1:1 mergers in $\sim$10$^{12}$, 5$\times$10$^{12}$, and
$\sim$10$^{13}$\msun halos, and 1:3 and 1:12 mergers with a $M_{\rm
  DM} \approx$ 10$^{13}$ \msun primary galaxy. We utilize 3 different
initial orbits for the galaxies.  For clarity, throughout this work,
we primarily focus on the results from a fiducial merger (in
Table~\ref{table:ICs}, model SMG10) which produces a relatively
average \sef$\approx$5-7 mJy SMG\footnote{When considering emission
  properties which are strongly dependent on merger mass, we will, at
  times, employ the usage of SMG1 as a ``high-mass'' fiducial SMG
  along with SMG10. This allows us to bracket the range of galaxy
  masses which form SMGs in our simulations. SMG1 forms an SMG ranging
  from average (\sef$\approx$5-7 mJy during inspiral) to extremely
  luminous and rare (\sef$\approx$15 mJy during final coalescence),
  whereas SMG10 forms at peak (final coalescence) only an average
  \sef$\approx$5-7 mJy SMG. See \citet{nar09} for more details
  regarding the lightcurve of model SMG1.}. The CO results (next
section) are generic for all models considered here. There is a
dispersion amongst the simulated submillimeter fluxes when varying
merger orbit (even at a constant galaxy mass; see Figure 2 of
\citet{nar09} for a direct quantification of this dispersion). The
fiducial SMGs studied here lie in the middle of this dispersion for
given halo masses, and, as such, are typical.  See \citet{nar09} for
results regarding the physical properties of these merger simulations.

\subsection{Radiative Transfer}
\subsubsection{Molecular Line Radiative Transfer}

We utilize the 3D non-LTE molecular line radiative transfer code,
\turtlebeach, to calculate the CO emission properties from our model
SMGs \citep{nar08a}. \turtlebeach \ is an exact line transfer code
that assumes full statistical equilibrium. The details of the code can
be found in \citet{nar06b} and \citet{nar08a}, and we refer the reader
to these papers for details on the specific algorithms used. Here, we
summarize the details relevant to this work.

\begin{figure}
\includegraphics[angle=90,scale=0.35]{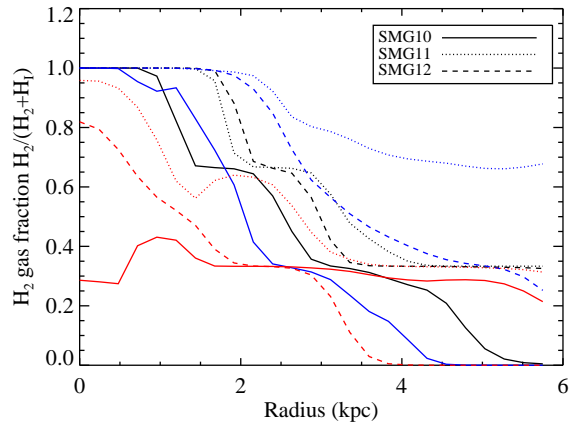}
\caption{Radially averaged molecular gas fraction as a function of
  galactocentric radius for three model SMGs (varying only the
  orientation angle of the disks) during three different phases. The
  three model SMGs produce ``average'' SMGs with fluxes ranging from
  \sef$\approx$2-7 mJy \citep{nar09}. The black lines are the
  pre-merger phase for each model; the blue lines denote the peak of
  the starburst (roughly the peak of the SMG phase); and the red lines
  represent the post-quasar phase. The molecular gas fraction is
  calculated via a dependency on the ambient pressure as motivated by
  observations of clouds in local galaxies \citep{bli06}. During the
  peak of the SMG phase (blue lines), the bulk of the neutral gas in
  the central regions is molecular, owing to rather high gas densities
  in the galactic nucleus.\label{figure:molfrac}}
\end{figure}

We assume that all cold, star-forming gas in the hydrodynamic simulations is
neutral. Following the observational constraints of \citet{bli06}, we
model the molecular fraction based on the ambient hydrostatic
pressure:
\begin{equation}
\label{equation:blitz}
R_{\rm mol} = n_{\rm H2}/n_{\rm HI} = \left[ \frac{P_{\rm ext}/k}{3.5
    \times 10^4}\right]^{0.92}
\end{equation}
where $k$ is Boltzmann's constant, and the external pressure P$_{\rm
  ext}$ is calculated via the fitting formula derived by
\citet{rob04}:
\begin{equation}
\rm log \ P = 0.05(\rm log \ n_H)^3 - 0.246 (\rm log \ n_H)^2 +
1.749(\rm log \ n_H)-10.6
\end{equation}
In Figure~\ref{figure:molfrac}, we show the radially averaged
molecular gas fraction as a function of galactocentric distance for
three model SMGs\footnote{We note that this feature of the code is an
  improvement over the algorithms described in
  \citet{nar08a}. Specifically, previous works using \turtlebeach \ 
  have assumed half the neutral gas (by mass) was molecular, consistent
  with average conditions in local galaxies \citep{ker03}. While these
  assumptions tying the molecular gas fraction to global observations
  may reproduce average molecular emission patterns in galaxies and
  AGN \citep[e.g.][]{nar06a,nar08b,nar08a,nar08c}, it does not provide
  spatially resolved information regarding the molecular content. By
  tying the molecular gas fraction to the ambient pressure as
  motivated by observations of GMCs \citep{bli06}, we are able to more
  accurately model the spatial distribution of the molecular gas. We
  note that even more sophisticated models for treating the neutral
  gas breakdown exist in the literature
  \citep[e.g.][]{pel06,rob07b}. However, utilizing methodologies such
  as these becomes computationally infeasible when considering the
  numbers of simulated galaxies and snapshots modeled in this work.}.

In order to investigate numerous snapshots at relatively high temporal
resolution (5 Myr) for a number of merger models, the hydrodynamic
simulations were smoothed to a spatial resolution of $\sim$ 160
pc. Resolution tests presented in \citet{nar08a} which investigated
the spectral and spatial invariance of these methods confirm that the
resultant CO morphologies, lines profiles, and excitation conditions
are convergent at these spatial resolutions. In order to properly
account for the molecular density gradients which exist within cells
of this volume (i.e. dense cloud cores and diffuse cloud atmospheres),
we further model the molecular gas following the sub-grid
prescriptions of \citet{nar08a}. Specifically, the molecular gas
within the $\sim$160 pc grid cells is assumed to reside in a mass
spectrum of GMCs which are modeled as singular isothermal spheres. The
mass spectrum of clouds follows a powerlaw with index $\alpha$=1.8 as
motivated by observations of local clouds \citep{bli07}, though tests
have shown that \turtlebeach \ results do not vary so long as the mass
spectrum indices reside within observational constraints
\citep{nar08b}. More details of the implementation of these sub-grid
treatments of GMCs may be found in \citet{nar08a}. Within the
molecular gas, we conservatively assume that Galactic abundance
patterns hold, and model the CO fractional abundance as CO/\htwo =
1.5$\times$10$^{-4}$. While molecular abundances in the ISM of high
redshift galaxies are unconstrained, recent measurements have shown
that star-forming UV-selected galaxies, likely progenitors of massive
mergers, have solar abundances in their ISM \citep{sha04}. As such, an
assumption of Galactic molecular abundances may be reasonable.

We build the emergent spectrum by integrating the equation of
radiative transfer through the \htwo \ gas:
\begin{equation}
  I_\nu = \sum_{r_0}^{r}S_\nu (r) \left [ 1-e^{-\tau_\nu(r)} \right
  ]e^{-\tau_\nu(\rm tot)}
\end{equation}
where $I_\nu$ is the frequency-dependent intensity, $S_\nu$ is the
source function, $r$ is the physical distance along the line of sight,
and $\tau$ is the optical depth. 

The source function is made up of a combination of the emission from
dust, as well as line emission. Formally, $S_\nu = j_\nu/\alpha_\nu$
where $j_\nu = j_\nu({\rm dust}) + j_\nu({\rm gas})$. Similarly,
$\alpha_\nu$ has components from both the gas and dust. The dust
radiates as a blackbody, and is assumed to be at the kinetic
temperature of the molecular gas. We note that this element of the
calculation is inconsistent with the \sunrise \ calculations, which
formally derives the dust temperature assuming the dust and radiation
field are in radiative equilibrium. Here, \citet{wei01} opacities are
assumed, though this makes little impact on the CO line flux.

The line source function is dependent on the CO level
populations. Therefore, in order to self-consistently calculate the
line intensities of CO, the molecular excitation properties must be
known. The relevant physical processes in determining the CO
excitation are collisions and radiative (de)excitation
\citep[e.g. line trapping;][]{nar08a}.  The molecules are assumed to
be in statistical equilibrium, and the population levels are
calculated considering both the radiation field and collisional
processes.

The methodology is an iterative one. To determine the solution to the
molecular excitation, the level populations across the galaxy are
first guessed at (in practice, we guess a solution near LTE). The
molecular gas is then allowed to radiate model photons based on the
assumed level populations, and, when a sufficient number of photons
have been realized in each grid cell, the mean intensity field is
calculated \citep{ber79}. Under the assumption of statistical
equilibrium, the radiative excitation rates in combination with the
collisional excitation rates give updated level populations, and new
model photons are then emitted. This process is repeated until the
level populations are converged. 

The line transfer takes into account velocity fields. The three
dimensional velocity field across the model galaxy is returned by the
\gadget \ hydrodynamic simulations. The line of sight velocity
gradient from cell to cell is accounted for in the line transfer via
emission and absorption line profiles, both of which are Gaussian in
nature. The emission and absorption profiles have their widths
determined by the thermal line width in the cell, as well as an
assumed microturbulent velocity field (here, set at 0.8 \kmsend). The
difference in the frequency centers of the emission and absorption
profiles is determined by the line of sight velocity difference
between the emitting and absorbing clump \citep[e.g. Equations 5 and 6
  of ][]{nar06b}.

For the models presented here, $\sim$13 million model photons were
emitted per iteration. The mass spectrum of GMCs are considered with a
lower cutoff of 1$\times$10$^{4}$ \msunend, and upper limit of
1$\times$10$^6$\msun \citep[consistent with constraints provided by
  local GMCs;][]{bli07}. The CO excitation was solved for across 10
levels at a time, and the collisional rate coefficients were taken
from the {\it Leiden Atomic and Molecular Database} \citep{sch05}. The
boundary conditions included the cosmic microwave background which was
modeled at \z=2.2 to have a temperature of T=8.74 K.

Finally, we comment that because of spatial resolution limitations in
the molecular line radiative transfer, we are forced to consider 8 kpc
boxes. Rather than following the center of mass of the merging galaxy
system (which, at times, may have scant little gas), we choose to
follow a single progenitor galaxy through its evolution. Of course, as
the nuclear disks of the progenitor galaxies overlap (both during
first passage, as well as coalescence), our models will include the
emission from both galaxies in the pair.

\subsubsection{Polychromatic SED Radiative Transfer}

In order to identify when our simulated galaxies would be selected as
submillimeter luminous sources, we simulate the ultraviolet (UV)
through submillimeter continuum photometry using the 3D adaptive grid
polychromatic Monte Carlo radiative transfer code, \sunrise
\ \citep{jon06a,jon06b,jon09}.  \sunrise \ calculates the transfer of
UV through millimeter wave radiation through the dusty interstellar
medium. We refer the reader to these work for details on the
underlying algorithms as well as numerical tests. Here, we summarize,
and explain the physical parameters employed in this study.

The radiative transfer is implemented via a Monte Carlo algorithm in
which photon packets representing many real polychromatic photons
propagate through the dusty interstellar medium, and undergo
scattering, absorption, and reemission.  Model cameras are placed
around the simulated galaxy to sample a range of viewing angles. The
emergent flux is determined by the number of photons that escape the
galaxy unhindered in a given camera's direction, as well as those
scattered into or reemitted by dust into the camera. For the purposes
of the calculations presented here, we used 8 cameras placed
isotropically around the model galaxy.

\sunrise \  is able to handle arbitrary geometries for the sources
and dust.  The input spectrum includes contributions from stars and
black holes (AGN). The AGN input spectrum utilizes the
luminosity-dependent templates of \citet{hop07} of unobscured
quasars. The normalization of the input spectrum is set by the total
bolometric luminosity of the central black hole, $L_{\rm AGN}$=$\eta
\dot{M_{\rm BH}}c^2$. Again, $\eta$ is assumed to be 10\%
(c.f. \S~\ref{section:hydro}).

The stellar input spectrum is calculated utilizing the stellar
populations code, STARBURST99 \citep{lei99,vaz05}, where the ages and
metallicities of the stars are taken from the hydrodynamic
simulations. We assume a Kroupa initial mass function (IMF)
\citep{kro02}, consistent with recent results from observations of
\zsim 2 star forming galaxies \citep{dav08, tac08, van08}. The
stellar particles initialized with the simulation are assumed to have
formed over a constant star formation history. In order to match the
star formation rate and stellar mass of the first snapshot of the
simulation, this corresponds to a SFH of $\sim$250 Myr.

The stellar clusters with ages less than 10 Myr are assumed to reside
in their nascent birthclouds.  \sunrise \ models the effects of
reddening of the stellar spectrum through these birthclouds utilizing
results from the photoionization code \mappings
\ \citep{gro04,gro08}. \mappings \ calculates the transfer of
continuum radiation and lines through the HII regions (which evolve as
a one-dimensional mass-loss bubbles; \citet{cas75}) and
photodissociation regions (PDRs) surrounding stellar clusters
\citep{gro08}. The effect of modeling the obscuration of stellar
clusters by HII regions and PDRs is a redistribution of emergent UV
light into the far infrared/submillimeter bands. The HII regions
absorb much of the ionizing UV flux and contribute heavily to the
hydrogen line emission from the ISM as well as hot-dust emission. The
time-averaged areal covering fraction by PDRs ($f_{\rm pdr}$) is
related to the PDR clearing timescale as $f_{\rm pdr}$ =
$exp(-t/t_{\rm clear})$, and is taken to be a free parameter
\citep[see ][]{gro08}. These PDRs absorb much of the non-ionizing UV
radiation field, and contribute to the emergent PAH and FIR
emission. Here, we assume an areal covering fraction of unity. A
covering fraction $f_{\rm pdr}$=1 translates to a cloud-clearing
timescale longer than the lifetimes of O and B stars \citep{gro08},
and the main consequence of reducing this clearing time scale is to
reduce the emergent submillimeter flux at the time of the starburst
\citep{nar09}. The cloud clearing time scales in gas rich galaxy
mergers are unconstrained. That said, some constraints may be placed
on this value either by educated guesses, as well as ansatzes that are
then verified by a comparison of the simulated SEDs to those observed.

The assumption of a clearing time scale longer than the lifetimes of O
and B stars may be a reasonable guess.  The centers of local gas rich
major mergers are known to have large molecular volume filling
fractions, and are well characterized by a uniform, smooth molecular
medium \citep{dow98,sak99} which may blanket nuclear O and B stars
their entire lives. While an assumption of $f_{\rm pdr}$=1 is not the
same as a uniform molecular medium, tests have shown that in this
limiting case, the submillimeter SED is the quite similar to $f_{\rm
  pdr}$=1 \citep{nar09}.

Similarly, we can take the reverse approach, and assume the ansatz of
$f_{\rm pdr}$=1, and compare the simulated SEDs to those observed. In
\citet{nar09}, we showed that the mean SED in modeled SMGs with a
range of masses compares quite well with observed SEDs of SMGs with
spectroscopic redshifts. Moreover, those authors found that the peak
submillimeter flux was related to the total mass of the
galaxy. Utilizing PDR clearing time scales longer than the lifetimes
of O and B stars resulted in average (\sef $\sim$ 5 mJy) SMGs with
halo masses of order $\sim 5 \times 10^{12}-10^{13}$ \msunend,
consistent with the inferred halo masses of observed SMGs
\citep{swi08}. Similarly, the simulated stellar masses, black hole
masses, and \htwo \ masses both compared well with observations, and
scaled with peak submillimeter flux. Reducing $t_{\rm clear}$ (or,
equivalently, $f_{\rm pdr}$) would require more massive galaxies to
produce relatively average \sef $\sim$ 5 mJy galaxies, and quickly
violate the inferred halo masses from clustering measurements. As
such, the assumption of $f_{\rm pdr} = 1$ may be reasonable. That
said, it is important to interpret the results presented in this paper
in the context of this assumption for the birthclouds surrounding
stellar clusters, especially in the context of mass-dependent results
(such as the CO line widths; \S~\ref{section:linewidths}).

The dust and radiation field are assumed to be in radiative
equilibrium, and utilize a method similar to that developed by
\citet{juv05}. Here, when photons are absorbed in a grid cell, and the
dust temperature updated, a new photon with SED equal to the
difference between the SED emitted for the new dust temperature and
that from the old one is emitted. This procedure is iterated upon
until the radiation field has converged \citep{jon09}.  The \sunrise
\ calculations employ 10 million photon packets per
iteration. \citet{jon09} find that the implementation of
dust-temperature iteration in \sunrise \ recovers the solution to the
\citet{pas04} radiative transfer benchmarks to within a few percent
for UV through mm wavelengths. The \citet{pas04} benchmarks were
recovered through the most stringent test cases of $\tau = 100$. We
note that the maximum optical depth we see during the SMG phase of
these simulations is $\tau \sim 75$.

 The gas and stars initialized with the simulation are assigned a
 metallicity according to a closed-box model such that Z=(-$y$
 \ ln[$f_{\rm gas}$]) where $Z$ is the metallicity, $y$ the
 yield=0.02, and $f_{\rm gas}$ is the initial gas fraction (though
 note the fluxes during the SMG phase of the model galaxies are not
 very sensitive to these assumptions).  Because the dust properties of
 \zsim 2 SMGs are relatively unconstrained, we utilize Galactic
 observations as input parameters. We assume a constant dust to gas
 ratio comparable to observations of local ULIRGs of 1/50. Tentative
 evidence suggests comparable dust to gas ratios in SMGs
 \citep{sol05,gre05,kov06,tac06}. Simulations parameterizing the dust
 in terms of a Galactic dust to metals ratio of 0.4 \citep{dwe98} give
 similar results to within 10\%. The dust grain model used is the
 R=3.1 \citet{wei01} dust model including updates by \citet{dra07}.

We are exploring the dust and photometric properties of \zsim 2
galaxies in a number of companion papers at various stages of
preparation. In this work, we focus on the CO properties of SMGs; as
such, we primarily utilize \sunrise \  as an informant as to when
the model galaxies may have sub-mm luminosities comparable to observed
SMGs. We classify our model galaxies as SMGs when they have an
observer-frame 850$\mu$m flux $\geq$ 5 mJy which is comparable to a
3-$\sigma$ detection in current wide-field surveys \citep[e.g.
][]{cop06}. We model our sources at \z = 2.5, and therefore this
fiducial criterion corresponds to a flux limit at a rest-frame
wavelength of $\lambda$=243 $\mu$m.

\section{Evolution of Sub-mm, $B$-band and \htwo \ properties of SMGs}
\label{section:evolution}

\begin{figure*}
\includegraphics[scale=0.9]{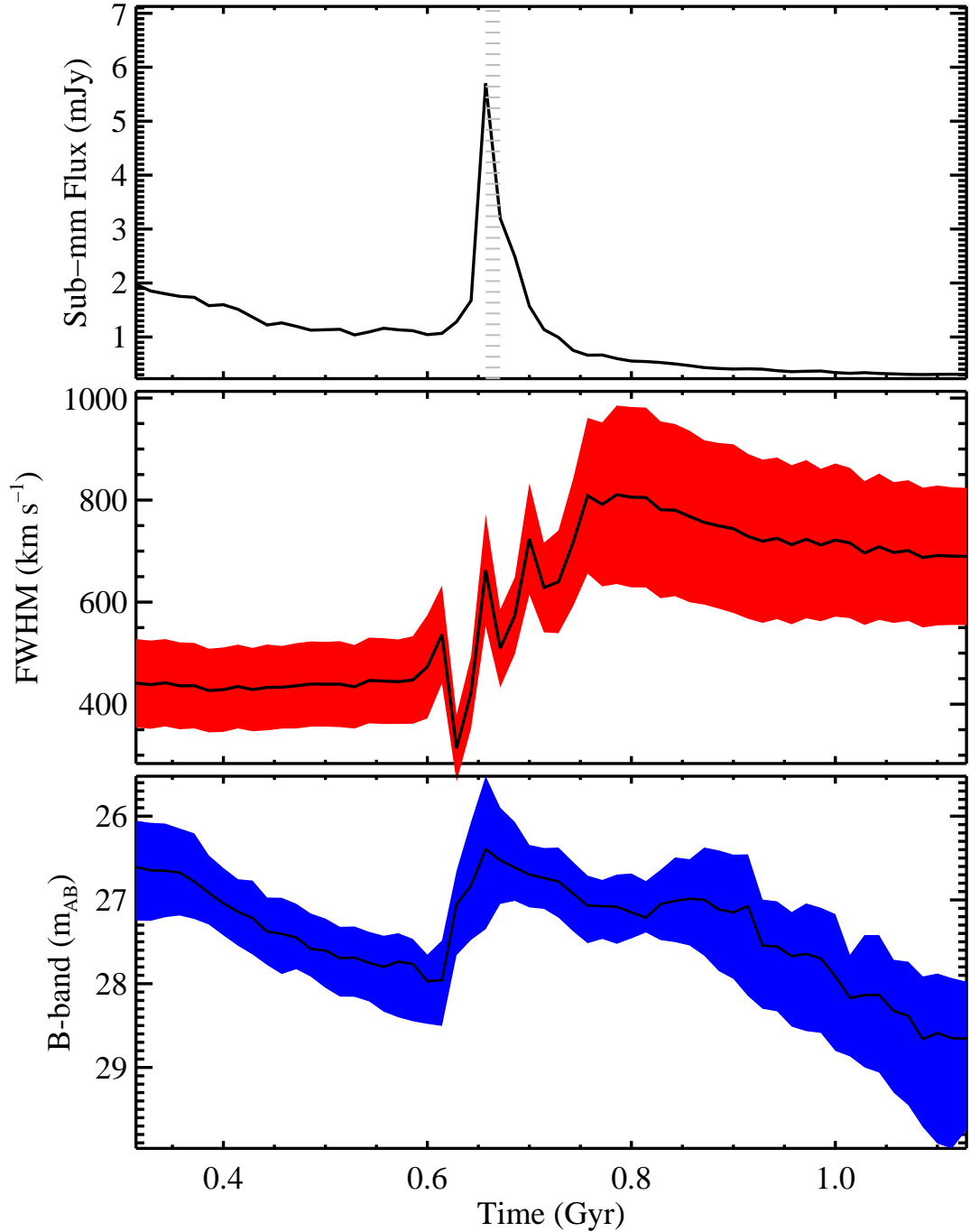}
\caption{Evolution of 850 \micron \ flux (top row), CO (J=3-2) line
  width (second row) and observed-frame $B$-band flux (AB magnitude;
  third row) for model galaxy SMG10 (Table~\ref{table:ICs}, designed
  to be an average SMG). All quantities are plotted at redshift
  \z=2.5. The grey hatched region in the top plot shows when the
  galaxy is visible as an SMG above \sef $>$ 5 mJy. The red shaded
  region in the CO line width plot shows the dispersion over 100
  random sightlines, and the blue shaded region in the $B$-band plot
  denotes the dispersion over 8 cameras placed isotropically around
  the galaxy. As the galaxies spiral in toward coalescence, the
  observed submillimeter flux is relatively low (\sef$\approx$1 mJy),
  and the progenitor galaxies are disk-like. Consequently, the line
  widths are representative of the virial velocity of the galaxies
  (here, set at $V_{\rm c}$=320 \kmsend).  As the galaxies coalesce
  ($T \approx$ 0.6-0.65 Gyr), the $\sim$1000 \msunyr starburst drives
  the 850 \micron \ flux to detectable levels (\sef $>$ 5
  mJy). Concomitantly, the $B$-band flux rises sharply, both owing to
  the intense starburst and a rapidly growing AGN. The CO line FWHM
  doubles as two disks enter the simulation box/observational beam
  (here, set at 8 kpc).  In the post-starburst phase, the
  submillimeter and $B$-band flux drops, and the line widths settle
  toward the rotational velocity of the combined (two-galaxy)
  system. See main text for more details. \label{figure:superplot}}
\end{figure*}

\begin{figure*}
\includegraphics[scale=0.9]{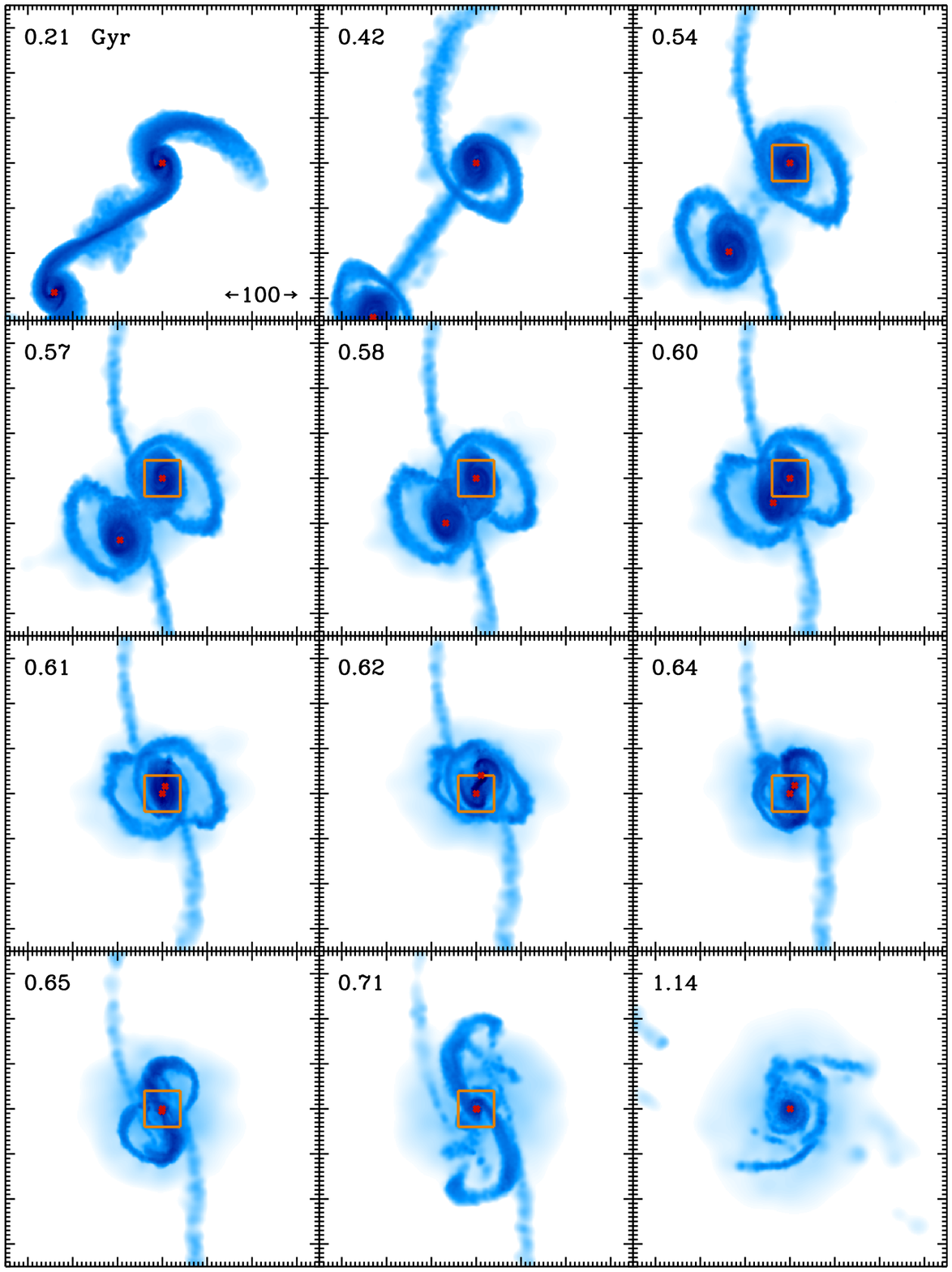}
\caption{Projected gas density for a fiducial merger model SMG10
  (designed to be an average SMG). Box sizes are 100 kpc on a side,
  and the box centers on the same galaxy the CO calculations center
  on. The time stamps are in the upper left of each panel, and in
  units of Gyr. For reference, the SMG phase is at time $T \approx$
  0.65 Gyr. Time stamps near the SMG phase have panels with equivalent
  time stamps to Figures~\ref{figure:smgmorph} and
  ~\ref{figure:vcent}. For these panels, the yellow box highlights the
  8 kpc simulation box employed for the CO radiative transfer
  calculations. The location of black holes is noted by the red
  crosses. \label{figure:panelfig}}
\end{figure*}


We begin with a general description of the evolution of the 850
\micron, $B$-band, and \htwo \ properties of our model SMGs, and
relate these to the evolutionary status of the galaxy merger.  In
Figure~\ref{figure:superplot}, we present the 850 \micron \ flux, CO
line width (which will be discussed more in
\S~\ref{section:linewidths}), and observed $B$-band apparent magnitude
\citep[AB magnitude system, modeled at \z=2.5, typical of SMGs;
][]{cha03a,cha05} of our fiducial merger simulation, SMG10
(Table~\ref{table:ICs}). The blue shaded region represents the
sightline dependent range of potential $B$-band magnitudes. The CO
transition plotted is J=3-2 in the rest-frame, corresponding to
mm-wave observations at \zsim 2. As a reference for the global
morphology, in Figure~\ref{figure:panelfig}, we plot the projected gas
density at various snapshots for fiducial model SMG10, with the
location of the black holes overlaid.

The initial passage of the galaxies induces a starburst of order
$\sim$200 \msunyrend. The galaxies form stars at this rate for a few
$\times$ 10$^{8}$ yr as they inspiral toward final coalescence. During
this time, the galaxy builds a stellar mass of order
$\sim$10$^{11}$\msun \citep{nar09}. This elevated SFR allows the
galaxy to produce $\sim$1-2 mJy of flux at 850 \micron \ rendering this
galaxy below the nominal 5 mJy detection threshold for SMGs, though
detectable with future sensitive instruments.

 When the galaxies approach toward final coalescence (T$\approx$0.6
 Gyr), tidal torques from the merger drive large-scale inflows
 \citep{bar91,bar96,mih94a,mih96}. Physically, the interaction of the
 galaxies spin up the disks as they transfer angular momentum from the
 orbit of the galaxies to the disks themselves, and triggers the
 growth of bars. Gas that shocks on this bar dissipates energy and
 loses angular momentum, causing an inflow to the central regions
 \citep[see, e.g. ][]{hop08d,hop09d}. The high densities in the
 nuclear region of the coalesced system give rise to a massive
 starburst of order $\sim$1000-1300 \msunyrend.  The nascent PDRs
 surrounding young stellar clusters convert the UV flux intercepted
 from O and B stars into longer wavelength submillimeter
 radiation. During this starburst event, the galaxy may be selected as
 a luminous submillimeter source with 850 \micron \ fluxes approaching
 $\sim$5-7 mJy \citep{nar09}.
 The peak submillimeter flux observed is directly related to the mass
 of the merging galaxies - galaxy mergers above total (halo) mass
 $\sim$5$\times$10$^{12}$\msun will produce the nominal $\sim$5 mJy at
 850 \micron \ to be detectable as an SMG. Mergers of a significantly
 lower mass will have difficulty producing a strong enough starburst
 to drive the observed submillimeter flux \citep{nar09}. This owes to
 the fact that the submillimeter flux in our model derives largely
 from reprocessing of UV flux from the starburst in the birth clouds
 surrounding stellar clusters \citep{gro08}.

Concomitant to the final coalescence starburst, inflows fuel central
black hole accretion.  A fraction of the accreted mass energy is
deposited into the ISM surrounding the central black hole(s), driving
a pressure-driven wind.  These AGN winds expel much of the obscuring
gas and dust in the central regions, allowing the system to be viewed
as an optically selected quasar \citep[T$\approx$0.65-0.7 Gyr; e.g.][
  and references therein]{hop05a,hop06,spr05a}. The quasar phase and
SMG phase are roughly coincident \citep[though the quasar phase may
  lag the SMG phase by up to $\sim$20 Myr;][]{spr05a}. This is in good
agreement with observations by \citet{cop08}, who find an overlap in a
subset of their \zsim 2 observed SMGs and quasars.  As demonstrated by
Figure~\ref{figure:superplot}, however, there is a large
sightline-dependent dispersion in potential $B$-band fluxes during the
final merger. Consequently, the same galaxy can show almost 2
magnitudes dispersion based on observed viewing angle, and not all
SMGs will appear as quasars. However, the galaxy is visible as an SMG
at all modeled sightlines.

It is important to note that the selectability of our simulated
galaxies as quasars is mass-dependent. As discussed by \citet{nar09},
the final black hole mass of the merged system is dependent on the
mass of the galaxy. This is similar to results found by \citet{lid06}
and \citet{li07} who found that quasar luminosity is tied to
progenitor galaxy halo mass. \citet{nar09} found that the final black
hole masses of average SMGs (e.g. \sef$\approx$5-7 mJy) will be a
few$\times$10$^{8}$ \msunend, whereas the mergers which produce the
most luminous SMGs (\sef $\approx$15-20 mJy) may make black holes
comparable to those seen in quasars ($M_{\rm BH}
\approx$10$^{9}$\msunend). We therefore continue the discussion of the
CO properties of our SMGs during the ``quasar phase'' as referring to
the time period when the simulated dust-attenuated $B$-band flux peaks
(e.g. third panel, Figure~\ref{figure:superplot}), and note that this
may not necessarily correspond to a galaxy selectable as a quasar in
current surveys. The exact relationship between SMGs and quasars will
be discussed in due course (J. Younger et al. in prep; D. Narayanan et
al. in prep).

The starbursts induced by the merger consume significant amounts of
gas. However, while gas consumption is high, supernova pressurization
of the ISM sustains large molecular gas fractions.  Thermal energy input
into the hot phase ISM, as well as mass increases owing to the stellar
mass returned and evaporation of cold clouds increases the ambient
pressure on cold clouds \citep{spr03a}. This increase in
pressure increases molecular fractions in the neutral ISM, in
accordance with our pressure-based \htwo \ formation/destruction
algorithm (c.f. Equation~\ref{equation:blitz}). Hence, while star formation
of course consumes \htwo \ gas, during starbursts, this effect may be
mitigated owing to conversion of HI to \htwo. The \htwo \ masses
during the final coalescence burst are rather high with typical masses
$\sim$5$\times$10$^{10}$\msunend, though can be as high as
$\sim$3$\times$10$^{11}$\msun \citep{nar09}.

\section{The Evolution of CO Properties of SMGs}

\subsection{CO Morphologies}
\label{section:morphology}
\subsubsection{Molecular Disk Formation and Disruption}
\label{section:diskformation}

\begin{figure*}
\includegraphics[scale=0.9]{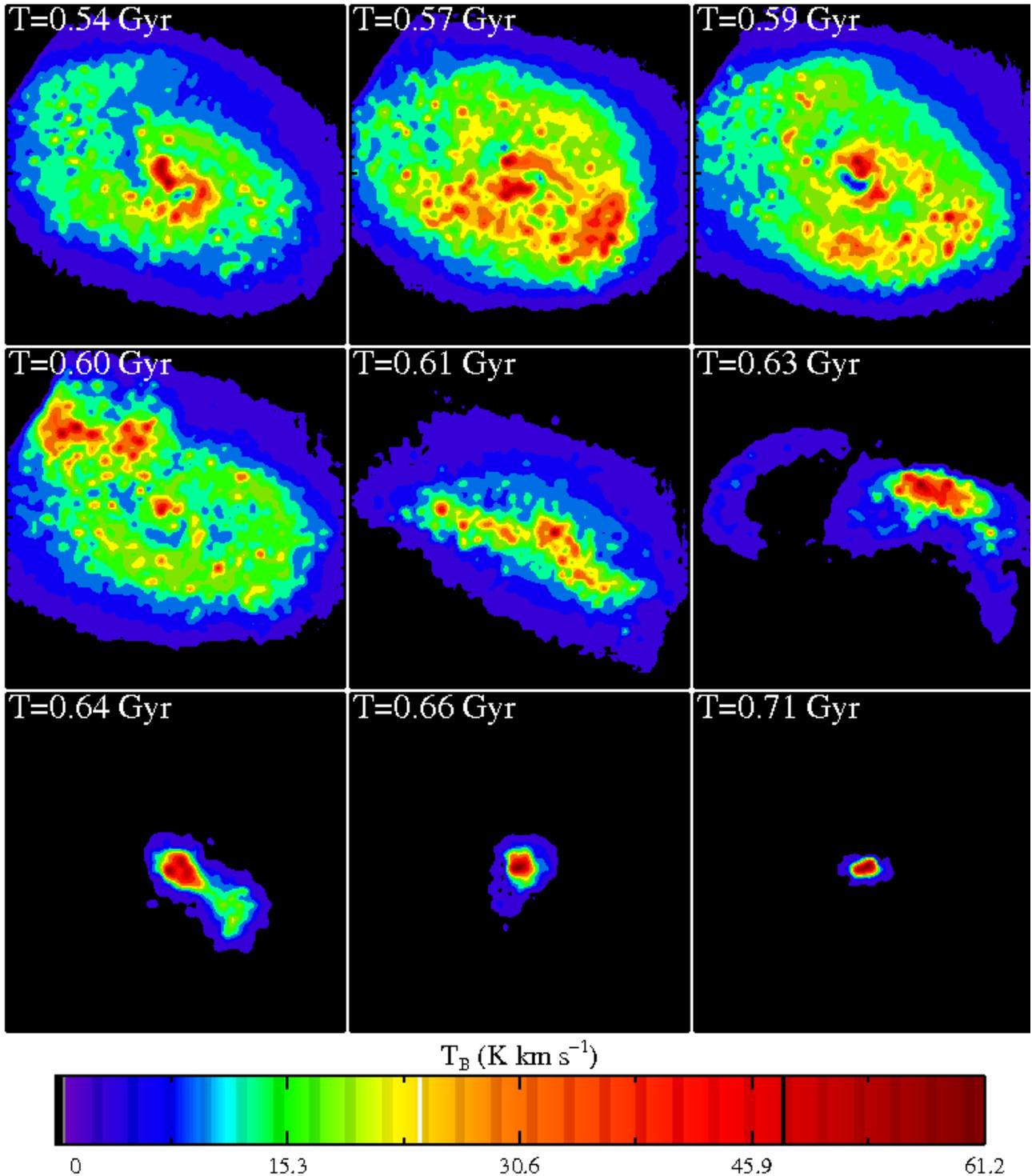}
\caption{Simulated CO (J=3-2) morphology for fiducial model SMG10. The
  boxes are 8 kpc on a side, and the intensity is in
  velocity-integrated brightness temperature with scale on bottom.
  The simulations focus on a single galaxy through its evolution until
  both galaxies are within a single 8 kpc box (second row and
  beyond). For reference, the 8 kpc box employed for the CO radiative
  transfer simulations is shown explicitly with respect to the global
  morphology in Figure~\ref{figure:panelfig}. During inspiral (first
  row of this figure) the molecular gas is in a disk-like
  configuration.  As the second galaxy enters the box, and they merge
  (second row), the disk-like morphology is disturbed and a large
  fraction of the gas is pulled from disk-like motion into relatively
  radial orbits. This corresponds to the peak of the SMG phase. During
  this time, extended features and tidal tails may become apparent
  (middle row). Tidal torquing drives much of the gas toward the
  central regions, resulting in a relatively concentrated molecular
  gas spatial extent (third row).  \label{figure:smgmorph}}
\end{figure*}

\begin{figure}
\includegraphics[angle=90,scale=0.35]{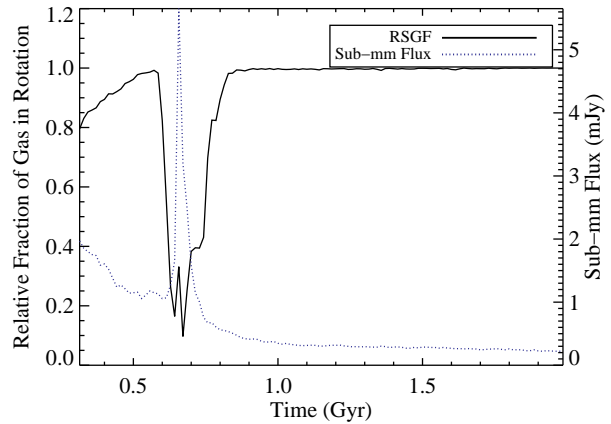}
\caption{Relative fraction of gas in rotational motion about the
  galactic nucleus (solid line, units on left axis), and 850 \micron
  \ flux (dotted line, units on right axis) for fiducial merger model
  SMG10. Following Cox et al. (2009, submitted), gas is considered to
  be in in rotational motion when it has circular velocity at least
  75\% of the expected Keplerian velocity at its radius.  During the
  peak of the SMG phase, most of the gas is disturbed from the central
  rotationally supported disk. \label{figure:submm_rsgf}}
\end{figure}

The structure of the molecular gas in SMGs, and in particular the
(potential) existence of molecular disks is essential to a thorough
understanding of the evolution of their CO line properties. We briefly
outline the key points related to this topic here, though defer a
detailed investigation into the survivability of molecular disks in
mergers to a future work.

In Figure~\ref{figure:smgmorph}, we show the CO (J=3-2) morphology of
galaxy SMG10 as a function of time. The temporal evolution depicts the
CO morphology as it evolves through inspiral (first row) and final
coalescence (second and third row).  The galaxy is most likely to be
viewed as an SMG during final coalescence when the SFRs are most
elevated \citep[e.g. Figure 1, ][]{nar09}. Here, this corresponds to
T$\approx$0.65 Gyr. Because our CO calculations center around a single
galaxy at all points in its evolution to maximize spatial resolution,
both galaxies only appear in the images when the nuclei are both
within the 8 kpc model box (second row and beyond).

In Figure~\ref{figure:submm_rsgf}, we show the relative fraction of
gas in rotational motion as a function of time, with the sub-mm flux
overlaid as a reference. Gas is considered to be in rotational motion
when its circular velocity exceeds 75\% of that expected at its radius
for Keplerian motion\footnote{Varying this fiducial fraction of 75\%
  makes no difference on the relative temporal evolution of the
  fraction of disk-like gas; lowering or increasing this value simply
  increases or lowers the normalization of the curve. As such,
  throughout this paper, this ``fraction'' should be taken as
  relative, and not absolute.}. As such, the relative fraction of gas
in rotational motion may be viewed as a measure of the ``diskiness''
of the molecular gas. Comparisons between
Figures~\ref{figure:smgmorph} and \ref{figure:submm_rsgf} may be made
via the time stamps displayed in the panels of
Figure~\ref{figure:smgmorph}.

\begin{figure*}
\includegraphics[scale=0.9]{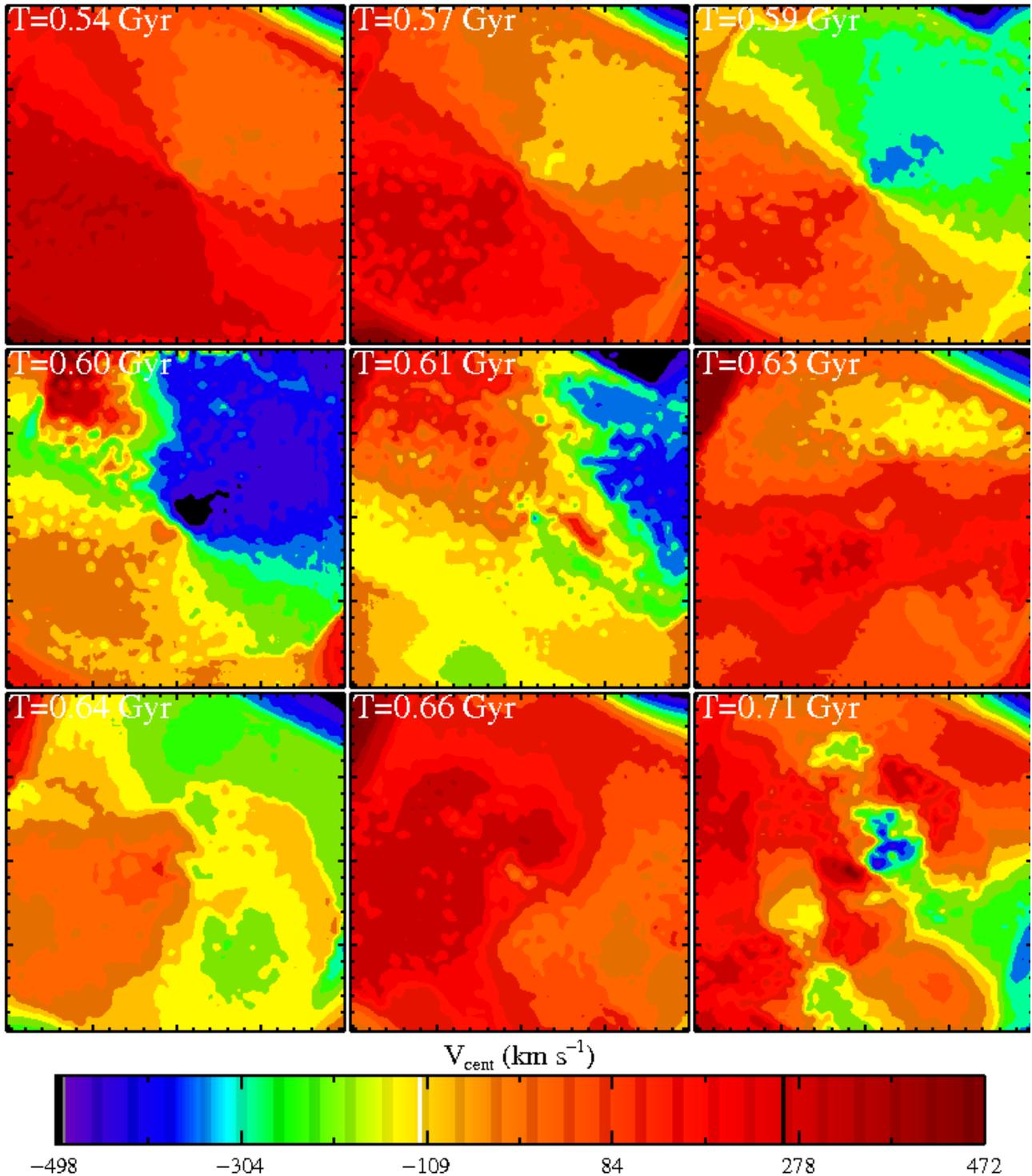}
\caption{Simulated CO (J=3-2) centroid velocity maps for model
  SMG10. The boxes are 8 kpc on a side, and the velocity are in units
  of \kms with scale on bottom. The snapshots match those of
  Figure~\ref{figure:smgmorph} The inspiral phase (top row) is
  characterized by ordered disk-like motion. The molecular disks are
  rapidly destroyed as the galaxies coalesce during the SMG phase ($T
  = 0.63-0.65$ Gyr). Note, the disk angle slowly changes throughout
  the early part of the galaxy's evolution, thus changing the
  magnitude of the line of sight velocities seen.\label{figure:vcent}}
\end{figure*}

The galaxies remain relatively disk-like as they inspiral toward final
coalescence.  Upon final merging, when the galaxy undergoes its
luminous SMG phase, the disk is tidally disturbed.  Tidal features
(and, on occasion, double-nuclei) become apparent in the molecular gas
morphology during this final-coalescence SMG phase (e.g. T$ \approx$ 0.65 Gyr,
Figure~\ref{figure:smgmorph}). This can be seen more explicitly in
Figure~\ref{figure:vcent}, where we plot the CO (J=3-2) centroid
velocity maps of the same snapshots shown in
Figure~\ref{figure:smgmorph}.  Soon after the final interaction/SMG
phase, the gas yet again re-virializes, and a strong (compact)
molecular disk re-forms (this is seen in
Figure~\ref{figure:submm_rsgf}, though the snapshots in
Figure~\ref{figure:vcent} do not extend far enough in time to show
this phase).  This history of molecular disk formation/disruption
throughout the galaxy merger's history is reminiscent of that seen in
models of the molecular ISM in \zsim 6 quasars \citep{nar08c}, and
will play an important role in our understanding of the CO line
widths, and usage of CO as a dynamical mass indicator in the sections
to come.

\subsubsection{Spatial Extent of CO Emission}
\label{section:halflight}
\begin{figure}
\includegraphics[scale=0.35, angle=90]{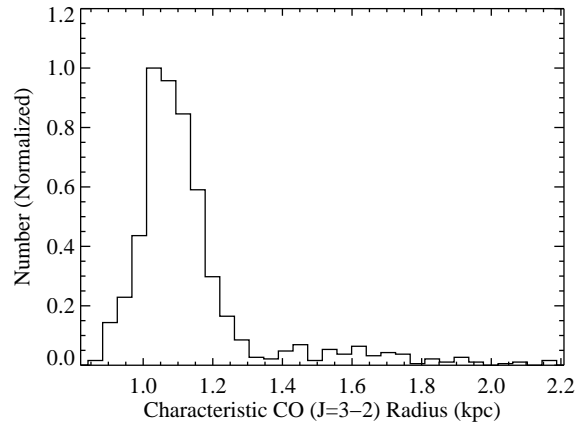}
\caption{Characteristic radius of CO (J=3-2) emission during the SMG
  phase of fiducial models SMG1 and SMG10. These models were chosen to
  bracket the range of masses which produce SMGs \citep[ranging from
    average, (5-7 mJy), to luminous ($\sim$15 mJy);][]{nar09}. The
  characteristic radius is calculated using the standard deviation in
  the flux-weighted radial distribution of fluxes. The distribution is
  plotted over 100 random sightlines.  The characteristic radii match
  up well with observed size scale FWHM measured by
  \citet{dow03,gen03,tac06,tac08} and \citet{you08b}. Please note,
  however, that because the simulations here are performed in
  isolation, rather than drawn from cosmological conditions, this
  distribution represents the range of expected CO (J=3-2) spatial
  extents from observed SMGs, rather than a true
  distribution. \label{figure:halflight}}
\end{figure}
With recent advances in (sub)mm instrumentation, high spatial
resolution images of the molecular gas in SMGs is beginning to become
available \citep[e.g.][]{tac08}. Constraints on the spatial extent of
the CO emission from SMGs can be important for e.g. determining the
dynamical mass from CO line widths.

During the merger-induced starburst, tidal torquing drives cold gas
from the outer disk into the nuclear regions of the
galaxy. Consequently, the CO emission becomes rather compact. In
Figure~\ref{figure:halflight}, we plot the distribution of
characteristic CO (J=3-2) radii from model galaxies SMG1 and SMG10
over 100 random sightlines during snapshots where the system may be
viewed as an SMG (\sef $>$ 5 mJy).  The characteristic radius for CO
emission is the flux-weighted standard deviation in the radial
distribution of fluxes. Practically, the map is treated as a histogram
of fluxes at varying radii. The standard deviation in this histogram
is the characteristic radius.

During the SMG phase, the emission is relatively compact owing to the
gas funneled into the central regions from the merger. The
characteristic radius averages at $\sim$1.5 kpc for most sightlines,
with some dispersion owing to evolutionary status and sightline. Some
large CO radii are seen from inspiralling disks in massive ($M_{\rm
  DM} \approx$ 10$^{13}$\msunend) mergers which are already SMGs
during the inspiral phase \citep{nar09}.

The CO spatial extents modeled here are comparable to the measurements
by \citet{tac06} of a $\sim$4 kpc FWHM diameter in SMGs (which
corresponds to a $\sim$0.85 kpc standard deviation in radius if the
emission is Gaussian in nature). The distribution of radii
additionally is consistent with the $\sigma \approx$ 0.6-3.3 kpc range
of spatial extents observed from SMGs
\citep{dow03,gen03,tac06,tac08,you08b,ion09}

\subsection{CO Line Widths}
\label{section:linewidths}



\subsubsection{Model Results: Evolution of CO Line Widths}

The observed molecular line widths of SMGs are exceptionally broad,
with a median FWHM of $\sim$600-800 \kmsend, and line widths exceeding
1000 \kms \citep{gre05,car06,tac06,cop08,ion09}.  Here, we explore the
evolution of CO line widths in our model SMGs; we show how our
merger-driven formalism for SMG formation and evolution may
self-consistently explain the observed broad lines from SMGs. Because
the CO emission line is essentially a distribution measuring the power
at a range of molecular gas line of sight velocities, rather than
employing any particular fitting methodology (and thus espousing the
associated uncertainties), we treat the line as a distribution of
fluxes, and utilize the standard deviation of the distribution
($\sigma$) as a measure of the line width. For a perfectly Gaussian
line, the FWHM of the line would simply be $\sim$2.35 $\times
\sigma$. While the lines are not perfectly Gaussian, to better compare
with observations, we utilize this conversion between $\sigma$ and
FWHM.

In short, the CO line widths are representative of the dynamics of the
system \citep{nar08c}.  Prior to final coalescence, typically only a
single galaxy is within the beam, and the line widths are narrow,
representing the rotational velocity of a single galaxy. During the
SMG phase, when the galaxies merge, the line widths roughly double
owing to the contribution of emission from both disks.  In the
post-SMG phase, as the molecular gas relaxes into a new disk
(Figure~\ref{figure:submm_rsgf}), the line widths drop by a factor
$\sqrt{2}$.

In more detail, the CO line widths reflect the dynamics of the
molecular gas in the galaxy. This owes to the origin of the CO
emission lines from the model galaxy. While emission within a given
cell (or neighboring cells) is typically optically thick, globally,
the emission is optically thin. When the CO line emitted from a cell
escapes the local region, it typically leaves the galaxy unhindered as
steep velocity gradients in the throes of the merger shift the
absorption profile out of resonance with the emission profile.  The CO
emission line, then, is essentially the sum of the emission that
escapes individual cells containing GMCs at their individual
velocities. As such, the line widths are reflective of the dynamics of
the system.  The dispersion ($\sigma$) in the CO line width
corresponds well with the velocity dispersion of the gas along the
line of sight.

As the galaxies inspiral toward final coalescence, the molecular gas
in the progenitor galaxies of the SMG is relatively virialized (Cox
et al., 2009, submitted), and thus the line widths are reflective of
the circular velocity of the progenitor disks. As discussed in
Table~\ref{table:ICs} and \citet{nar09}, our model mergers which
produce average SMGs (\sef$\approx$5-7 mJy) were typically initialized
with disks with $V_{\rm c}$=320 \kmsend, thus resulting in a CO
linewidth of $\sigma \approx$ 160 \kms (which is equivalent to the 320
\kms circular velocity in the disk modulated by sin(30$\degr$) to
account for the average inclination angle of the disk). This
corresponds to a FWHM of $\sim$375 \kms for a Gaussian velocity
dispersion (Figure~\ref{figure:superplot}). Recall that during the
inspiral phase, typically only one galaxy is within the 8 kpc beam,
which is comparable to most interferometric beam sizes
\citep{gre05,tac06}.

When the galaxies coalesce, both galaxies contribute to the detected
line. The velocity dispersion of gas detected doubles, and
consequently, the line width roughly doubles. This is coincident with
the SMG phase. Here, this causes the line widths to increase to 2
$\times$ $V_{\rm c}$. Because the FWHM $\approx$ 2.35 $\times V_{\rm
  c}$, and accounting for the average inclination angle of the disks,
we then arrive at a generalized expression for the CO line FWHM during
the final coalescence SMG phase:
\begin{equation}
\label{eq:fwhm}
{\rm FWHM_{SMG}} \approx 2 \times {\rm sin}(i) \times 2.35
  \times V_{\rm c}
\end{equation}
which is, of course, simply FWHM $\approx$ 2.35 $\times V_{\rm c}$
when an average inclination angle of $i$=30$\degr$ is assumed.  Here,
where circular velocities of $V_{\rm c}$=320 \kms are employed for the
initialization of the progenitor galaxies, this results in modeled
linewidths of $\sim$600-800 \kmsend.  The dispersion in the modeled CO
line widths owes to both viewing angle and evolutionary effects.

Generically, the SMG phase and quasar phase occur at or around the
nuclear coalescence of the galaxies. However, the exact timing of this
event is dependent on a number of things, including galaxy orbit, gas
content, and mass, amongst other factors. As such, the CO line widths
from the SMGs or quasar host galaxies with a merger-origin show a
large range of line widths.  In Figure~\ref{figure:fwhm_hist}, we plot
the histogram of CO linewidths from our fiducial galaxy (SMG10) during
both its SMG phase as well as its quasar phase. As before, we
nominally define the SMG phase as when the 850 \micron \ flux is $\ga$
5 mJy, and arbitrarily define the quasar phase as when the galaxy is
brighter than 27th magnitude (AB Magnitude; apparent magnitude at
\z=2.5). Because the SMG phase and quasar phase are nearly coincident,
the linewidths are generally broad throughout both phases, with nearly
indistinguishable distributions.

 \subsubsection{Observational Comparisons}

\begin{figure}
\includegraphics[angle=90,scale=0.35]{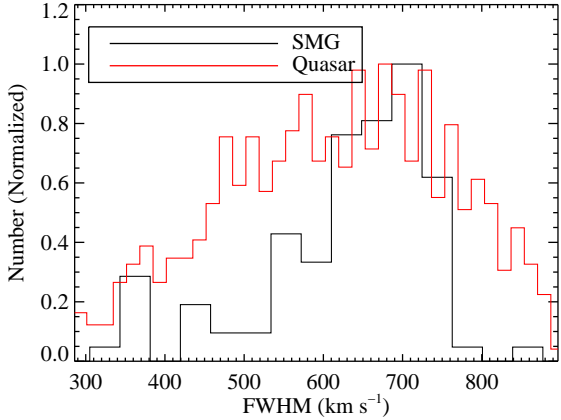}
\caption{Distribution of modeled CO (J=3-2) linewidths during the SMG
  and ``quasar'' phases of model SMG10. The SMG phase is considered
  when \sef$>$5 mJy, and the quasar phase is arbitrarily assigned with
  an apparent $B$-band magnitude (AB system) cut of 27th
  magnitude. See Figure~\ref{figure:superplot} for more details.  The
  phases occur at similar time periods (typically separated by at most
  $\sim$20 Myr), and consequently have similar line width
  distributions.  The FWHM distribution for these two sources appear
  to correspond well with the distributions published by
  \citet{cop08}, though the quasar distribution appears to be
  discrepant with those published by \citet{car06}. See text for
  details regarding the origin of the broad CO line widths in SMGs and
  quasars. Again, because the simulations here are not cosmological in
  nature, the distribution should be viewed as representative of the
  range of CO line widths from average SMGs, not the true distribution
  from a blind survey. While we normally include higher mass models
  (e.g. SMG1) in our analysis, we refrain here as the extremely broad
  (FWHM$\approx$1000-1400 \kmsend) lines from SMG1 take away from the
  main point of modeled SMGs with average \sef \ fluxes producing
  average CO line widths. SMGs as massive as SMG1 will be rare
  (indeed, this is apparent by the relative lack of observed \sef$>$15
  mJy SMGs), though when observed, will have line widths FWHM $>$ 1000
  \kms on average. \label{figure:fwhm_hist}}
\end{figure}

The distribution of modeled line widths for SMGs shown in
Figure~\ref{figure:fwhm_hist} is in excellent agreement with those
presented by \citet{cop08} and \citet{car06}.  The linewidths are
broadly reflective of the mass of the simulated SMGs, and likely
signifies a correspondence between the final masses and evolutionary
status of our modeled SMGs and those in nature. In this sense, our
model faithfully provides a natural explanation for the broad line
widths observed in SMGs.

A comparison between the line widths of our model quasars and those in
nature is more difficult. \citet{cop08} presented new CO detections of
submillimeter-luminous quasars, and found a nearly indistinguishable
distribution of CO line widths from quasars and SMGs, in excellent
agreement with the simulations presented here. In both our models, and
the observations of \citet{cop08}, the median CO line width is
$\sim$600-800 \kmsend, which we view as a general success of our
model. However, a literature compilation of line widths from
CO-detected quasars by \citet{car06} found a much narrower median in
the distribution, with the line widths showing a median value of
FWHM$\approx$300 \kmsend, in contrast to both the observational
results of \citet{cop08} as well as the model results presented
here. The source of this is discrepancy not entirely clear. Because
the analysis performed by \cite{car06} was from a literature
compilation, the sample utilized a variety of CO transitions from a
non-uniform sample with objects spanning a large range of redshifts
from \zsim 1-6. Different CO lines trace different spatial extents,
and thus may show a range of line widths. Similarly, quasars even of a
similar mass as those presented in \citet{cop08} but at a lower
redshift would naturally show evolution in their line widths owing to
shallower gravitational potentials. In contrast, the observations by
Coppin et al. utilized only CO (J=2-1) and (J=3-2) emission from a
sample of quasars in a smaller redshift range (\z=1.7-2.6). In this
sense, the comparison between our models and the relatively uniform
observations of \zsim 2 quasars by Coppin et al. seems most
appropriate as it best compares to the redshifts and emission lines
investigated in this work.

\section{The Usage of CO as a Dynamical Mass Tracer in SMGs}
\label{section:mdyn}

\begin{figure}
\includegraphics[scale=0.35,angle=90]{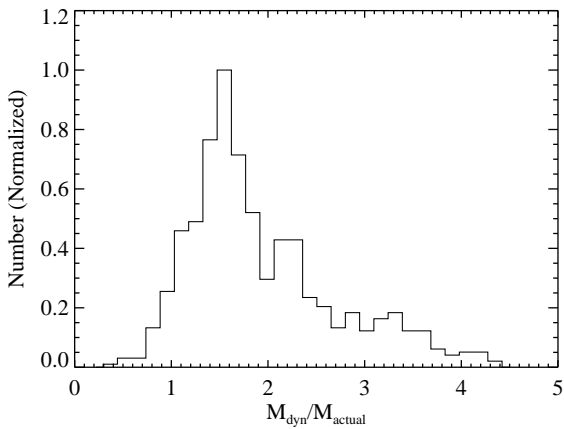}
\caption{ Sightline averaged distribution of \mdyn/M$_{\rm actual}$
  for models SMG1 and SMG10. These models were chosen to bracket the
  range of masses of galaxies which appear to produce SMGs in our
  simulations.  The dynamical masses are calculated as in
  Equation~\ref{eq:mdyn} for the simulated CO (J=3-2)
  emission. Generally, using CO line widths from SMGs for a dynamical
  mass calculation requires a modulation of a factor of $\sim$1.5-2
  for translating to a true enclosed mass. See text for
  details.\label{figure:mdyn}}
\end{figure}

In \S~\ref{section:linewidths}, we saw that during the SMG phase of
the model galaxy's evolution, the CO line widths were larger than
expected from ordered disk-like motion gas. In this context, it is
interesting to quantify the usage of CO as a dynamical mass tracer in
SMGs.

The exact value of CO line widths in high redshift mergers as a
dynamical mass indicator depends, of course, on the relationship
between linewidth and dynamical mass assumed. Broadly, for disk-like
motion, this relationship can be expressed in the form:
\begin{equation}
\label{eq:mdyn}
M_{\rm dyn} = k \frac{\Delta V_{\rm fwhm}^2 R_{\rm hwhm}}{G \ \rm {sin}^2(i)}
\end{equation}
where $\sigma$ is the 1D velocity dispersion in the line, $i$ the disk
inclination angle, $R$ the spatial extent of the CO emission (here,
taken to be the half-width at half-maximum), and $k$ a constant
encompassing the relationship between $V_{\rm c}$ and $V_{\rm fwhm}$
(for a given distribution of mass) and $R_{\rm g}$ and $R_{\rm
  hwhm}$. Equation~\ref{eq:mdyn} holds due to the global optical
thinness of molecular line emission across a galaxy.

However, uncertainties exist in all of these conversion
factors. Furthermore, it is not clear what fraction of the molecular
disk survives during the merger, and whether the inclusion of a
sin$^2$($i$) term is appropriate. Even if it were, prior to the high
resolution imaging capable only by ALMA, we can at best assume an
average disk angle of $i$=30$\degr$. As such, a constructive method
for quantifying the usage of CO as a dynamical mass tracer in SMGs is
to say:
\begin{equation}
\label{eq:mdyn}
M_{\rm dyn} = \frac{\Delta V_{\rm fwhm}^2 R_{\rm hwhm}}{G} = k' \times
M_{\rm actual}
\end{equation}
and characterize the relationship between $\Delta V_{\rm fwhm}^2
R_{\rm hwhm}/G$ and $M_{\rm actual}$. Here, note that $M_{\rm actual}$
is the total (baryonic and dark matter) mass enclosed in $R_{\rm
  hwhm}$, and $\Delta V_{\rm fwhm}$ is calculated as 2.35$\times$ the
standard deviation in the line width
(c.f. \S~\ref{section:linewidths}).  In Figure~\ref{figure:mdyn}, we
plot the ratio of the dynamical mass as calculated in
Equation~\ref{eq:mdyn} to the true enclosed mass (which includes dark
matter and baryonic mass). The radius employed is the half-width at
half maximum from the simulated CO emission maps. The ratio of $M_{\rm
  dyn}$/$M_{\rm actual}$ is calculated for SMGs which span the range
of masses explored here, for all snapshots which satisfy the fiducial
criteria \sef$>$5 mJy, and for 100 random sightlines. On average, the
ratio between the dynamical mass inferred from Equation~\ref{eq:mdyn}
and the true enclosed mass in the half-mass radius, $k'$,
ranges\footnote{Because these simulations are not cosmological, it is
  not possible to quote an exact mean in the distribution of $M_{\rm
    dyn}$/$M_{\rm actual}$. This range signifies the range seen when
  varying the exact model SMG or number of SMGs used to create a
  distribution similar to that plotted in Figure~\ref{figure:mdyn}.}
from 1.5-1.9. With this, it is then feasible to modulate
Equation~\ref{eq:mdyn} with this given value of $k'$ to construct the
appropriate relationship between CO-derived dynamical mass and true
mass enclosed in the CO emitting region.

\section{CO Excitation and Line Spectral Energy Distribution} 
\label{section:excitation}

The excitation of CO is an important diagnostic of high redshift
galaxies. First, molecular line measurements from SMGs are typically
made in millimeter bands, which corresponds to high excitation CO
lines in the rest frame at \zsim 2. Because the inferred molecular gas
mass is generally derived via converting CO (J=1-0) velocity
integrated intensity, understanding the typical CO line ratios (and
average level of thermalization of the gas mass within the telescope
beam) is crucial. Assumptions of thermalized line ratios (e.g. LTE,
when brightness temperature ratios between levels are unity) between
higher lying lines and CO (J=1-0) may underestimate the molecular gas
mass in the event of substantial quantities of subthermal gas. Second,
the CO excitation patterns reveal the line(s) of dominant CO power
output. As more broadband molecular line spectrometers become
available (e.g. the ZEUS, Z-spec, Zpectrometer spectrometers;
\citet{hai08,bra04,har07}), CO detections of high redshift galaxies
will be critically dependent on observations of the brightest
lines. In an effort to aid interpretation of existing data sets, and
help guide future observations of SMGs, we investigate the molecular
excitation properties and CO line ratios from our model SMGs. We do
not attempt to address the long standing problem of converting CO
(J=1-0) flux to \htwo \ gas mass in galaxies, but rather simply relate
the flux density from higher excitation lines to that from the ground
rotational transition.

\begin{figure*}
\includegraphics[scale=0.7,angle=90]{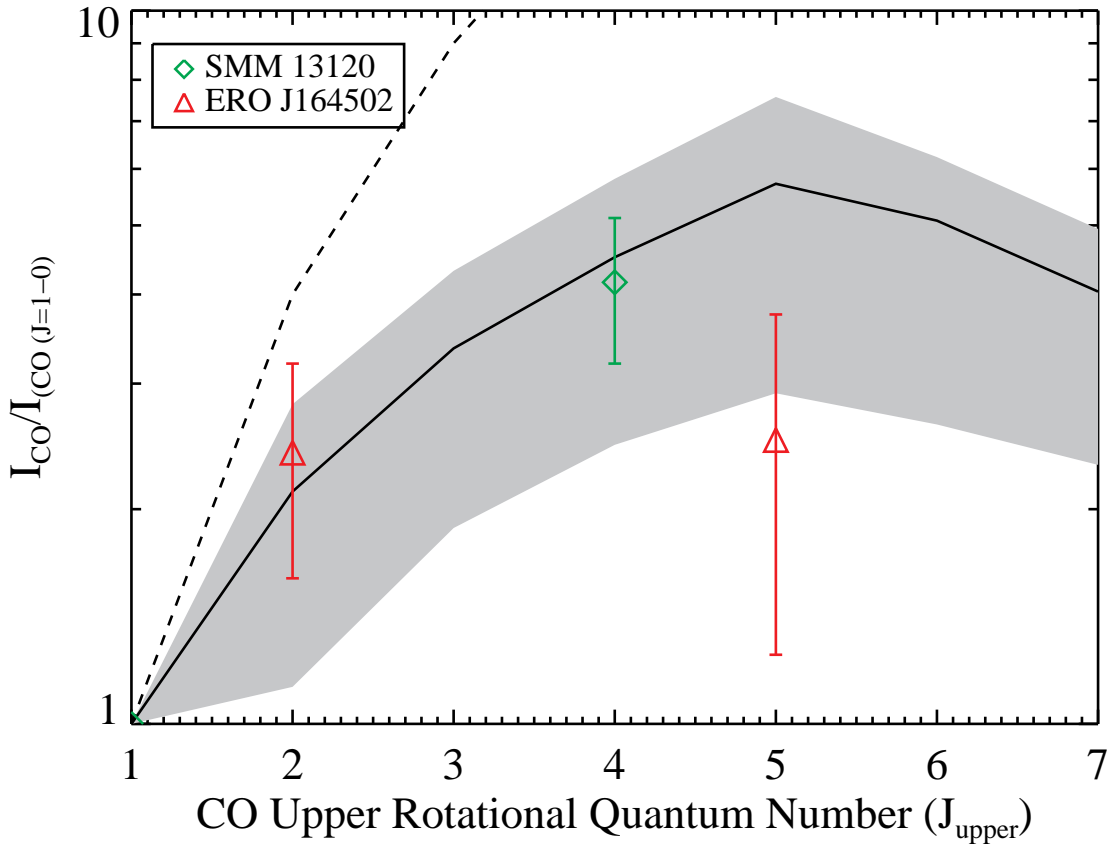}
\caption{Model CO SED from fiducial models SMG1 and SMG10. The SED is
  presented for all snapshots where the galaxy would be detected as an
  SMG (\sef $>$ 5 mJy), and the solid line shows the mean SED while
  the grey shaded region denotes the 1-$\sigma$ dispersion amongst
  snapshots. The flux densities from J$>$1 levels are compared against
  the CO (J=1-0) level.  The dashed line represents the predicted CO
  SED for thermalized level populations (LTE). The green diamond shows
  observational data from SMG SMM 13120+4242 \citep{hai06}, and the
  red triangles from SMG ERO J164502 (aka HR 10)
  \citep{and00,gre03,pap02}. To our knowledge, these represent the
  only SMGs with tabulated multi-line data (including a CO (J=1-0)
  detection) which are unlensed. The highest excitation CO SEDs come
  from final coalescence mergers, whereas massive SMGs that are above
  \sef$>$5 mJy during inspiral may have lower CO excitation. The mean
  SED displays reasonable agreement with the observations, and suggest
  that CO levels J $\ga$ 2 may not be thermalized. This shows that
  caution must be exercised when assuming a brightness temperature
  ratio of unity to the CO (J=1-0) line when deriving molecular gas
  masses. \label{figure:cosed}}
\end{figure*}

\subsection{Model Results: Highly Excited CO in SMGs}

In Figure~\ref{figure:cosed}, we plot the sightline-averaged model CO
SED for models SMG1 and SMG10 (Table~\ref{table:ICs}) which bracket
the mass range of galaxies that satisfy the nominal selection criteria
\sef $>$ 5 mJy. The ordinate is the ratio of the velocity-integrated
intensity from various transitions compared to CO (J=1-0), normalized
to CO (J=1-0). The range of emergent CO SEDs from our models is
denoted in the grey shaded region, and the mean by the solid line. The
dashed line shows the expected CO SED for thermalized level
populations.  The line ratios are modeled for unresolved observations,
and include emission from the entire 8 kpc simulation box. To best
compare with the few published constraints (next section), we plot the
line ratios relative to the ground (J=1-0) transition. This has the
added benefit of providing a direct measure of the ability of higher
lying transitions to convert to \htwo \ gas masses.

As Figure~\ref{figure:cosed} shows, there is a broad dispersion in
potential CO SEDs from our simulated SMG. However, two generic
features are evident. First, on average, the CO is quite excited with
the SED turnover only occurring at the J$\approx$5-6 level. This is in
reasonable agreement with observations of \zsim 2 SMGs (see next
section). The higher excitation CO SEDs come from SMGs which arise
during final coalescence mergers, while the lower excitation SEDs
represent the inspiral phase of massive galaxies (e.g. SMG1) which may
be seen as SMGs even prior to final coalescence \citep[e.g. Figure 1,
][]{nar09}.

Second, while the turnover is typically at a relatively high J level,
most CO levels are subthermal over the 8 kpc simulated box. The
rotational ladder in Figure~\ref{figure:cosed} show that the gas is
nearly thermalized for the lowest lying lines (CO J=2-1). Higher lying
lines, however, are not thermalized over the 8 kpc model box. While
the gas associated with the nuclear starburst in the central $\sim$1 kpc
is warm, dense, and thermalized, the outer regions (R$\sim$1-4 kpc)
contain a significant quantity of lower density gas which contributes
to the emergent spectrum via line trapping. This lowers the mean
excitation conditions observed. As such, Figure~\ref{figure:cosed}
demonstrates that most observations of SMGs which do not focus solely
on the nucleus will reveal subthermal CO emission at higher-lying
(J$\ga$3) levels\footnote{Higher spatial resolution observations
  probing the nuclei of SMGs should show higher excitation
  conditions. This lends itself to a direct testable prediction from
  these models which we outline in
  \S~\ref{section:observational_tests}.}. This is effectively saying
the filling factor of dense gas is relatively low, thus lowering the
mean observed CO SED. {\it Consequently, assumptions of thermalized CO
  line ratios from unresolved observations of SMGs will typically
  underpredict the molecular gas mass}.


\subsection{Observational Comparisons}
While few tabulated observational constraints exist for relative
intensities from SMGs, the excitation seen in
Figure~\ref{figure:cosed} seems to display reasonable agreement with
the existing published data. Line ratios for ERO J164502 and ERO
J164502 are given by \citet{and00,pap02,gre03} and
\citet{hai06}. These data are represented by the points with error
bars in Figure~\ref{figure:cosed}. While the CO SED is normalized to
the ground state (thus making this point an unconstraining
comparison), the higher excitation detection of two lines (J=2-1 and
J=4-3) is entirely consistent with the predictions made here. The
intensity from the J=5-4 line from ERO J164502 is lower than the
models predict here, though we note that this is lowest excitation SMG
known \citep{pap02,wei07}.

A detailed survey and compilation of literature data has been
presented by \citet{wei07}. While these data are not tabulated, we can
make qualitative comparisons. The bulk of these galaxies show
subthermal emission in the higher lying lines with a turnover at the
J=5 or 6 level. Thus, at face value, these observed CO SEDs are
comparable with the predictions made in Figure~\ref{figure:cosed}. 

A detailed examination of the Wei\ss \ et al. observed CO SEDs,
however, suggests a potential discrepancy between the observed values
and those modeled here. The observed intensity at the J=5 or 6 level
(where the SED turns over) is a factor of $\sim$15 above that from the
ground (J=1-0) transition; this is in contrast to our models which
suggest that the most excited lines will be only a factor of $\sim$5
greater than the ground state (Figure~\ref{figure:cosed}). This
difference may be reconciled with better constraints on the true CO
(J=1-0) emission from SMGs. Recall that the CO SED is typically
plotted (both here, as well as in observational literature) as
relative to the CO (J=1-0) transition. For the bulk of the observed
sources in the literature, the J=1-0 line has not directly observed,
but rather inferred from large velocity gradient (LVG) modeling which
is simplified approximate method compared to the 3D non-LTE radiative
transfer methods employed here. It may be that the apparent difference
in excitation between the observations and theoretical predictions
will be reconciled via future detections of CO (J=1-0) from SMGs.



\section{Discussion: Testable Predictions}
\label{section:discussion}
\subsection{Testable Predictions}
\label{section:observational_tests}

In this paper, we have outlined a model for the CO emission from high
redshift SMGs. Our models have provided a natural explanation for the
observed \htwo \ gas masses, CO spatial extents, line widths and
excitation conditions. We have attempted to compare with literature
data when available, and generally found a reasonable correspondence
between our model results and galaxies in nature.

This all said, while our models appear to provide a plausible match to
observed   data,  the   comparisons  made   thus far  are   in  essence
postdictions.  An even  more powerful  test of  any  theoretical model
would  be  testable  {\it  predictions}  of future  surveys.  In  this
section,  we sketch  out potential  observable tests  of  these models
which are imminently possible  with the latest generation of bolometer
and heterodyne receiver technology.
\subsection{CO Linewidths of \zsim 2 SMGs and Quasars}
\begin{figure*}
\includegraphics[angle=90,scale=0.25]{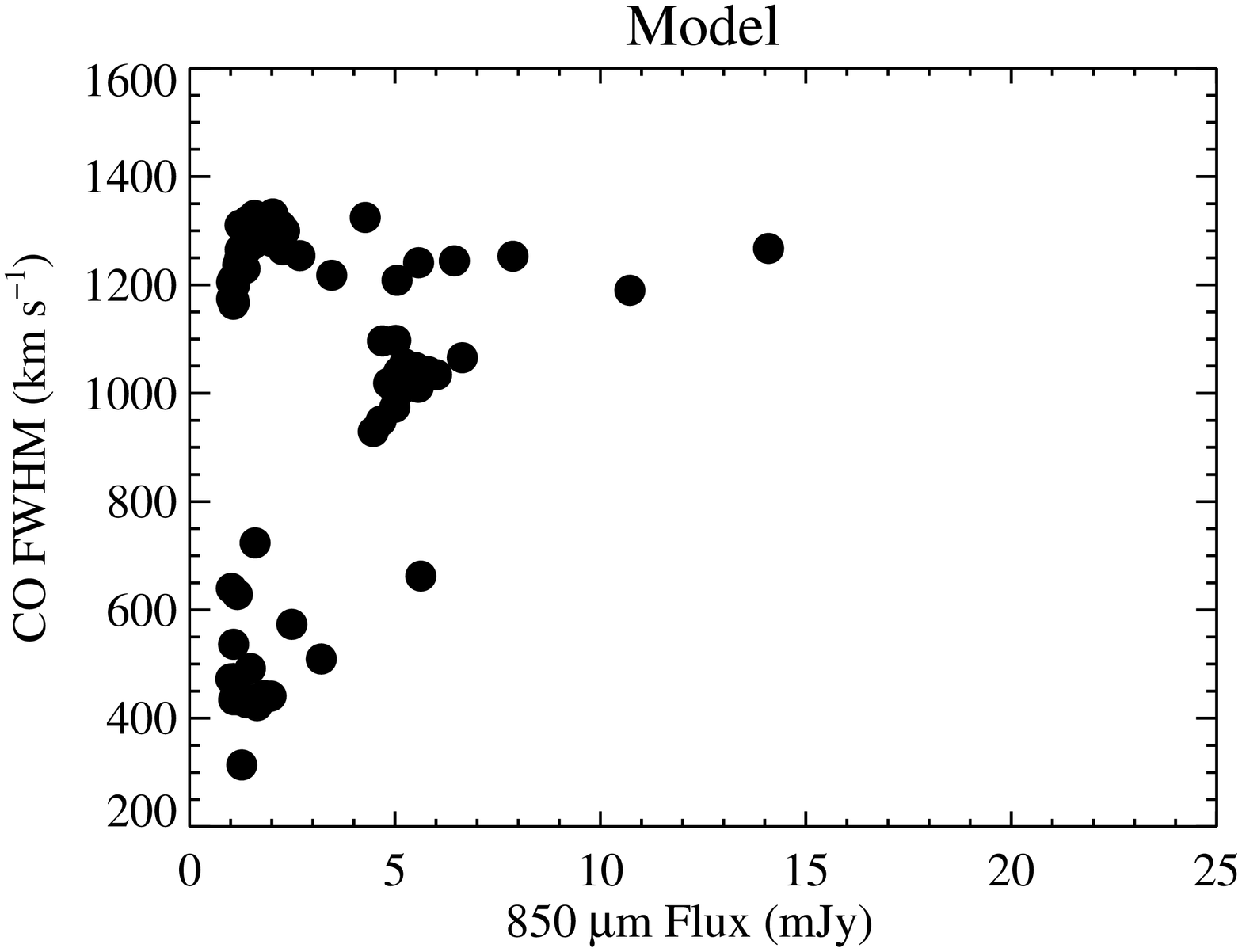}
\includegraphics[angle=90,scale=0.25]{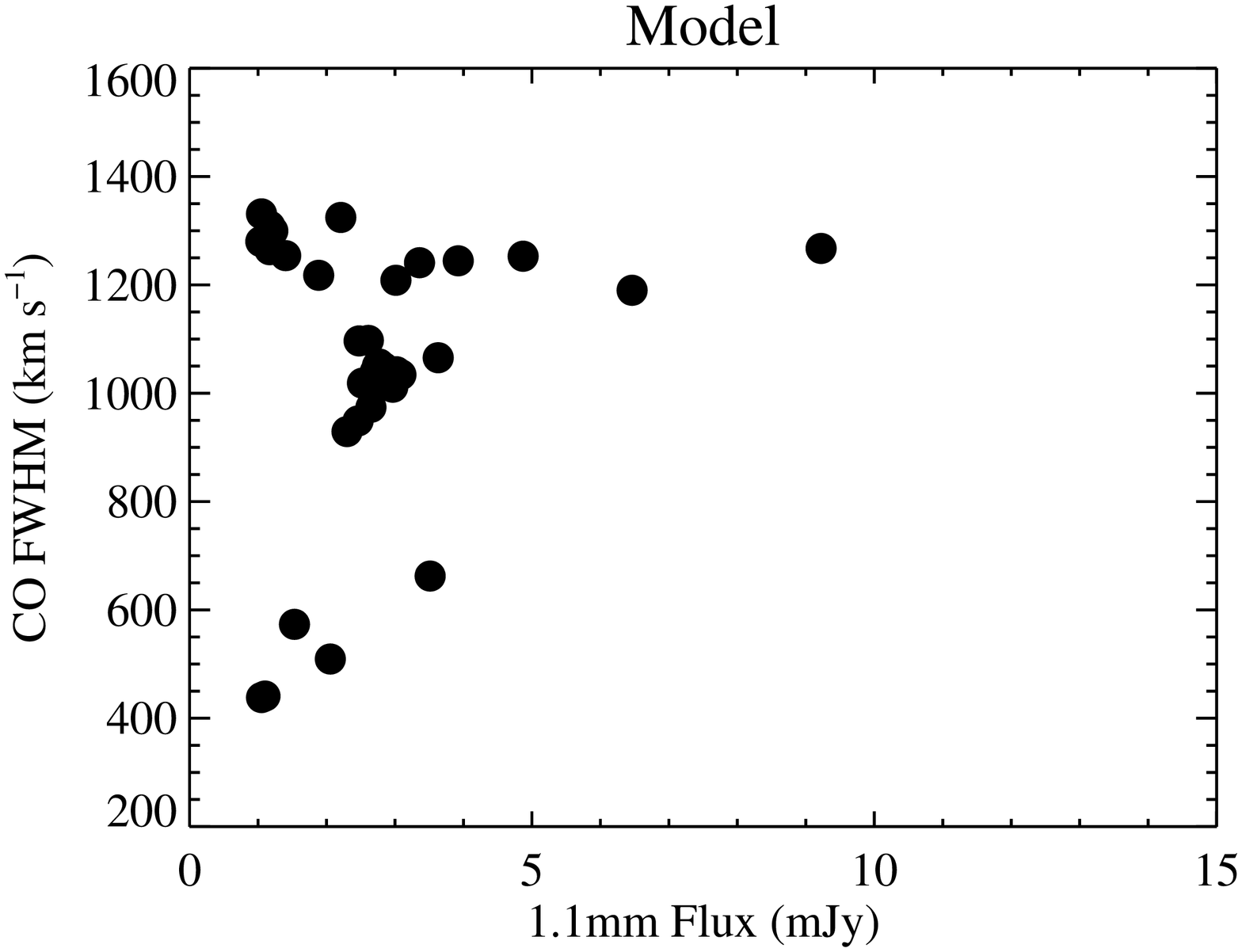}
\includegraphics[angle=90,scale=0.25]{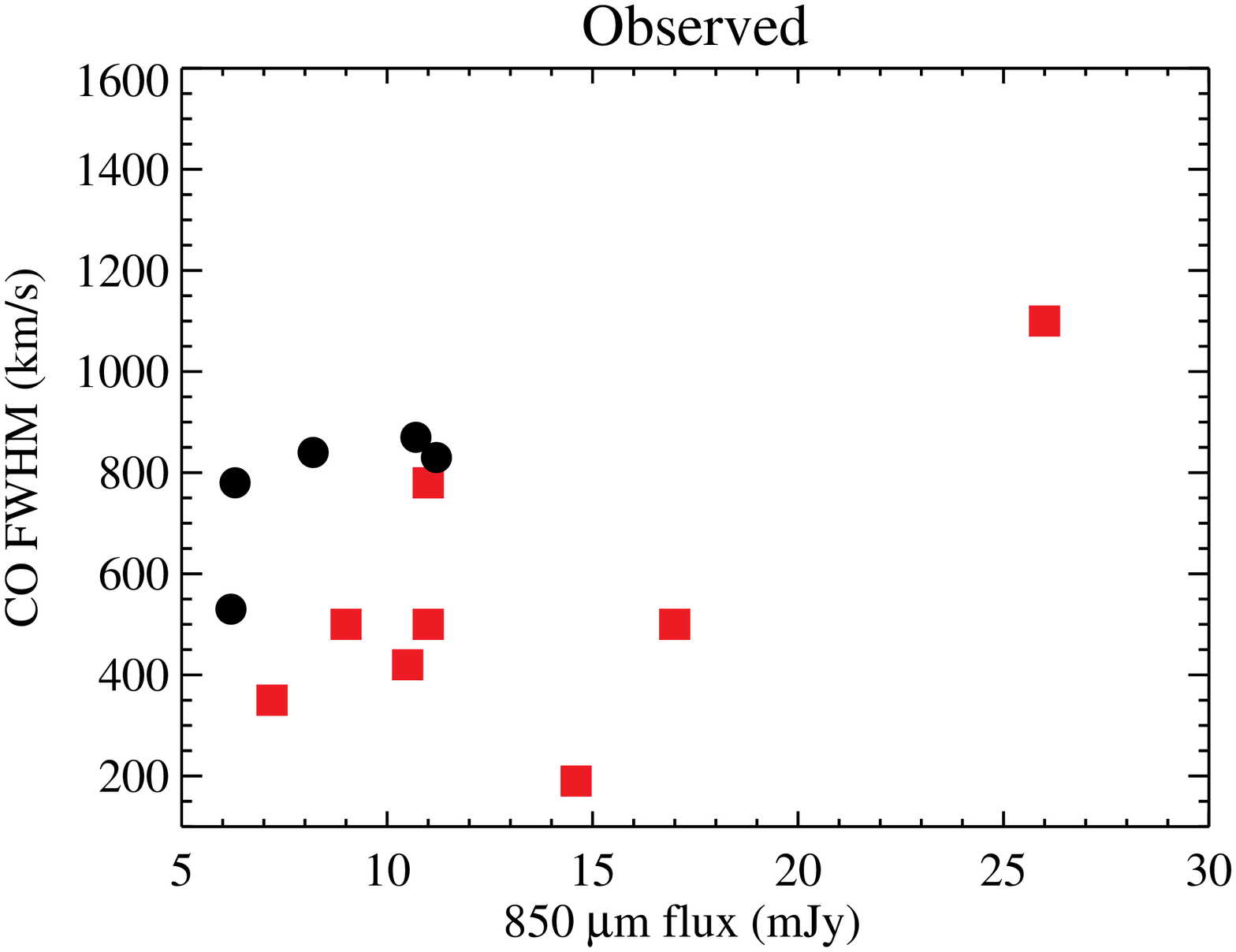}
\caption{Predicted CO (J=3-2) FWHM-\sef \ relation (left panel),
  predicted CO (J=3-2) FWHM-$S_{\rm 1.1 mm}$ relation (middle panel),
  and observed FWHM-\sef \ relation culled from data from
  \citet{sol05}. The models are for models SMG1, SMG10 and SMG13. SMG1
  and SMG10 were chosen as usual to bracket the range of halo masses
  employed here, and SMG13 added to probe lower fluxes (though note it
  is too low in mass to ever form a \sef $>$ 5 mJy SMG). The brightest
  SMGs at 850 \micron \ and 1.1 mm are predicted to have a narrow
  dispersion of CO FWHMs, and typically broad lines whereas lower
  luminosity objects may have a larger range of line widths. The mean
  CO line width will increase with increasing (sub)mm flux. See text
  for reasoning. From the observational dataset, optical quasars have
  been eliminated. The comparison to the observations is generally
  inconclusive.  This may owe to many uncertain gravitational lens
  magnifications in the observed SMGs \citep[known lensed sources are
    marked as red squares; ][]{sol05}.  Because these models probe
  both the full dynamic range of simulated 850 \micron \ fluxes and CO
  FWHMs, adding more models does not increase the dispersion in the
  models plots. \label{figure:fwhm_submmflux}}
\end{figure*}

Our model for the increased line widths from SMGs relies on a
temporary increase in velocity dispersions owing to multiple galaxies
in the simulation box/telescope beam.  Two features of this model are
evident. First, the line widths are relatively low, and representative
of the virial velocity of a single progenitor galaxy during the
inspiral phase when the observed 850 \micron \ flux may be relatively
low (\sef $\approx$ 1 mJy; Figure~\ref{figure:superplot}). They
increase concomitant to the increase in submillimeter flux, and are
thus broader when the galaxy is in its transient SMG phase, though
remain broad in the post-SMG phase when the galaxies 850 \micron
\ flux decreases again. In this picture, one might expect a broad
range of linewidths from galaxies with lower ($\sim$1 mJy) 850 \micron
\ fluxes: small line widths corresponding to inspiralling galaxies,
and large line widths corresponding to post-SMG phase galaxies.
Similarly, lower mass mergers (e.g. model SMG13) at final coalescence
may contribute to the dispersion of line widths on the low flux end.

Second, as discussed by \citet{nar09}, the 850 \micron \ lightcurve
from merging galaxies has a similar shape for a mass sequence of
mergers, though scales in normalization. This means that the
submillimeter flux curve shown in the top panel of
Figure~\ref{figure:superplot} simply scales upward with increasing
merger mass \citep[indeed this is evident with a direct comparison of
  the lightcurve presented in Figure~\ref{figure:superplot} with that
  of Figure 1 in ][]{nar09}. This means that in more massive mergers,
the inspiral phase corresponds to 5 mJy sources, and the peak SMG
phase to rarer, more luminous (\sef$\ga$10-15 mJy) SMGs. Consequently,
there may be a spread in CO line widths from even 5 mJy SMGs. The most
luminous sources, however, only occur at the final coalescence burst
of extremely rare $\sim$10$^{13}$\msun mergers. As such, \sef
$\approx$ 10-20 mJy galaxies will have extremely broad (FWHM $>$ 1000
\kmsend) CO line widths\footnote{Recall that the CO line width scales
  with circular velocity, and consequently, with galaxy mass.} with a
relatively small dispersion.

 The aforementioned effects will have the following generic
 consequence for CO linewidths from high redshift mergers: the mean CO
 FWHM will increase with increasing submillimeter flux, and the
 dispersion will decrease.  We show this explicitly in
 Figure~\ref{figure:fwhm_submmflux} by plotting the sightline-averaged
 model CO FWHM versus 850 \micron \ flux for model galaxies SMG1,
 SMG10 and SMG13\footnote{These models were chosen to span a broad
   mass range. Note that model SMG13 is too low in mass to form a
   detectable \sef$>$5 mJy SMG; \citet{nar09}.} in
 Table~\ref{table:ICs} for a flux limit of \sef$>$1 mJy, and in the
 middle panel, the CO FWHM-1.1 mm relation. This serves as a
 prediction for both future sensitive 850 \micron \ surveys, as well
 as 1.1 mm \citep[e.g. AzTEC;][]{wil08a} counts.

In the right panel of Figure~\ref{figure:fwhm_submmflux}, we attempt
to compare our model prediction in the left panel with data culled
from the recent review by \citet{sol05}. An important point is that
optically selected quasars cannot be included in this sort of
comparison as they may preferentially have their molecular disks in a
face-on configuration, thus skewing average line widths
\citep[e.g.][]{car06,nar08c}. This does exclude a number of sources
which may be simultaneously quasars, and submillimeter bright
\citep[e.g. Figure~\ref{figure:superplot} and ][]{cop08}. The observed
data in the right panel of Figure~\ref{figure:fwhm_submmflux} are
generally inconclusive in comparison to the predictions in the left
panel. The effects of uncertain gravitational lens magnifications
muddy interpretation \citep[sources that are known to be lensed are
  marked as red squares; ][]{sol05}. Better statistics with SCUBA2 and
AzTEC will provide a direct and imminent test for this aspect of these
models.

\subsection{Predicted CO Excitation for High-Resolution Observations of SMGs}
\begin{figure}
\includegraphics[angle=90,scale=0.35]{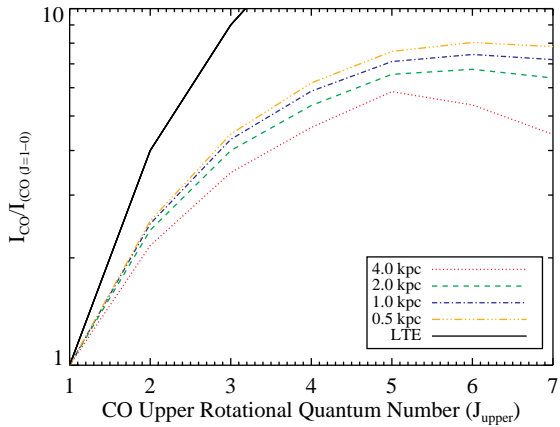}
\caption{Predicted CO line SEDs for models SMG1 and SMG10 during their
  SMG phases at increasing spatial resolution. The black solid line
  shows the predicted CO line SED for thermalized populations. As the
  warmer, denser gas toward the nucleus is probed, the peak and shape
  of the excitation ladder shift toward higher levels as more of the
  gas becomes thermalized. High resolution observations by ALMA of the
  nucleus of SMGs will typically show higher excitation CO line SEDs
  than lower resolution observations. \label{figure:cosed_nuc}}
\end{figure}

With the advent of broadband receivers on a multitude of telescopes,
CO SEDs of high redshift objects will soon become available and serve
as a test for the excitation predictions outlined in
\S~\ref{section:excitation}.

We first note that Figure~\ref{figure:cosed} itself is a testable
prediction. As few multi-line surveys are currently published,
Figure~\ref{figure:cosed} serves as a direct comparative for future
constraints on the CO SED from SMGs. A crucial component to this will
be the direct detection of CO (J=1-0) emission from SMGs. In the
absence of this, relative intensities to this transition will be
difficult to interpret, and direct conversion of emission from higher
lying transitions to \htwo \ masses unclear. This is currently
feasible at a variety of telescope facilities, though interferometers
(e.g. the EVLA) may be preferable as baselines may be problematic on
single dish telescopes \citep{hai06}.

Second, we may make predictions for imminent higher spatial resolution
observations (e.g. ALMA). The CO excitation conditions presented in
Figure~\ref{figure:cosed} for our model SMGs were averaged over the
entire 8 kpc box, thus including emission from subthermally excited
gas. While the gas associated with the nuclear starburst is indeed
warm and thermalized, including diffuse, subthermal gas had the effect
of lowering the mean excitation condition. This is analogous to
observations of \zsim 2 SMGs which include both dense, nuclear gas, as
well as diffuse gas in the telescope beam. Consequently, the modeled
excitation conditions in Figure~\ref{figure:cosed} showed excellent
agreement with multi-line CO measurements from SMGs.

In principle, one can imagine that higher spatial resolution
observations of SMGs which focus on the dense, nuclear regions would
show higher excitation CO SEDs. Our approach allows us to quantify
this effect, as well as make predictions for the next generation of
high spatial resolution interferometers (e.g. ALMA). In
Figure~\ref{figure:cosed_nuc}, we plot the sightline-averaged CO SED
for SMG models SMG1 and SMG10 as a function of decreasing physical
beam size. As before, we only consider SMGs with a fiducial selection
criteria of \sef $\ga$ 5 mJy. As the observations probe narrower and
narrower columns, the observed gas is on average more highly excited,
and the turnover point of the CO SED moves to increasingly high
rotational number. The lack of complete thermalization in the lower
observed transitions (e.g. CO J=2-1) owes to the fact that diffuse gas
is still folded into the observation, even for rather high spatial
resolution observations.

\section{Summary}
\label{section:summary}

We have combined hydrodynamic simulations of Submillimeter Galaxy
formation and evolution with 3D non-LTE molecular line radiative
transfer calculations to provide a model for the CO emission
properties from SMGs. Our model has shown a number of successes in
matching the observed spatial extent of CO emission, CO line widths
and excitation conditions. We utilized these models to understand the
origin of these emission properties:

\begin{itemize}
\item In our model, SMGs originate in major mergers
  \citep{nar09}. Strong gaseous inflows drive highly concentrated
  molecular gas complexes such that the observed characteristic CO
  radius is of order $\sim$1.5 kpc. The large radius tail of the
  distribution arises from pre-coalescence galaxies in extremely
  massive ($\sim$10$^{13}$\msun) halos which are SMGs even during the
  inspiral phase \citep[e.g. Figure 1 of ][]{nar09}.

\item The large CO line widths from SMGs owe to the fact that they are
  typically being observed during a transient phase where the gas is
  highly non-virialized and multiple galaxies are in the simulation
  box/telescope beam. During interactions, the CO FWHM from mergers is
  roughly 2.35$\times$ the circular velocity of a progenitor (for a
  Gaussian line; Equation~\ref{eq:fwhm}). Two merging $V_{\rm c}
  \approx$ 320 \kms disks naturally produce average (\sef$\approx$5-7
  mJy) SMGs with extremely broad line widths of order $\sim$600-800
  \kmsend.

\end{itemize}

We have additionally been able to provide interpretation regarding the
usage of CO as a diagnostic of physical conditions:

\begin{itemize}

\item The usage of CO line widths from SMGs as a dynamical mass
  estimator may overestimate the enclosed mass. Typical overestimates
  are of order \mdyn/ M$_{\rm actual} \approx$1.5-2.

\item The CO excitation in SMGs is high, with the rotational ladder
  turning over at the $\sim$J=5 or 6 level. The level populations at
  J$_{\rm upper}>$ 2 are typically subthermal, and assumptions regarding brightness
  temperature ratios of unity with the ground state will lead to
  underestimates of the inferred \htwo \ gas mass. The CO (J=3-2) line
  from SMGs is typically a factor of $\sim$3 below the intensity
  expected from thermalized level populations.

\end{itemize}

Finally, we have made predictions for this model which are imminently
testable with the newest generation of bolometer arrays and wide-band
sensitive CO receivers:

\begin{itemize}

\item The CO line widths from galaxies which will become SMGs are
  predicted to be their broadest when the submillimeter flux is the
  highest. Consequently, the brightest SMGs (\sef $>$ 15 mJy) are
  predicted to only have quite broad (FWHM$>$ 1000 \kmsend) line
  widths with a small dispersion in FWHMs observed. Lower flux
  galaxies (\sef $\approx$1) can be either normal disks (where the
  line widths are predicted to be relatively narrow), or post-SMG
  phase mergers (where the line widths will be relatively broad). As
  such, lower flux galaxies are predicted to have a broad dispersion
  in observed CO FWHM. Care must be taken to both remove optically
  selected quasars (which may be biased to have face-on molecular
  disks) and lens-magnified sources from the sample.

\item The intensity from the CO J=5-4 or 6-5 line, where the CO SED
  turns over, is predicted to be a factor of $\sim$5-7 times that of
  the ground state. While detections at these higher lying lines are
  becoming routine, future detections of CO (J=1-0) emission from SMGs
  will be required to test the predicted rotational ladders.

\item The peak and shape of the CO SED will shift toward higher lying
  transitions as smaller spatial extents are investigated with future
  high resolution arrays (e.g. ALMA).

\end{itemize}

\section*{Acknowledgements} We thank Kristen Coppin, Reinhard
Genzel, Thomas Greve, Laura Hainline, and Daisuke Iono for helpful
discussions and comments on an earlier draft of this paper. We are
extremely grateful to Patrik Jonsson and Brent Groves for their
extensive help with \sunrise. The simulations in this paper were
run on the Odyssey cluster supported by the Harvard FAS Research
Computing Group.  This work was partially funded by a grant from the
W.M. Keck Foundation (TJC) and a NSF Graduate Student Research
Fellowship (CCH).

\bibliographystyle{apj}
\bibliography{/Users/dnarayanan/paper/refs}

\begin{thebibliography}{109}
\expandafter\ifx\csname natexlab\endcsname\relax\def\natexlab#1{#1}\fi

\bibitem[{{Alexander} {et~al.}(2005{\natexlab{a}}){Alexander}, {Bauer},
  {Chapman}, {Smail}, {Blain}, {Brandt}, \& {Ivison}}]{ale05a}
{Alexander}, D.~M., {Bauer}, F.~E., {Chapman}, S.~C., {Smail}, I., {Blain},
  A.~W., {Brandt}, W.~N., \& {Ivison}, R.~J. 2005{\natexlab{a}}, \apj, 632, 736

\bibitem[{{Alexander} {et~al.}(2005{\natexlab{b}}){Alexander}, {Smail},
  {Bauer}, {Chapman}, {Blain}, {Brandt}, \& {Ivison}}]{ale05b}
{Alexander}, D.~M., {Smail}, I., {Bauer}, F.~E., {Chapman}, S.~C., {Blain},
  A.~W., {Brandt}, W.~N., \& {Ivison}, R.~J. 2005{\natexlab{b}}, \nat, 434, 738

\bibitem[{{Alexander} {et~al.}(2008)}]{ale08}
{Alexander}, D.~M. {et~al.} 2008, \aj, 135, 1968

\bibitem[{{Andreani} {et~al.}(2000){Andreani}, {Cimatti}, {Loinard}, \&
  {R{\"o}ttgering}}]{and00}
{Andreani}, P., {Cimatti}, A., {Loinard}, L., \& {R{\"o}ttgering}, H. 2000,
  \aap, 354, L1

\bibitem[{{Barger} {et~al.}(1998){Barger}, {Cowie}, {Sanders}, {Fulton},
  {Taniguchi}, {Sato}, {Kawara}, \& {Okuda}}]{bar98}
{Barger}, A.~J., {Cowie}, L.~L., {Sanders}, D.~B., {Fulton}, E., {Taniguchi},
  Y., {Sato}, Y., {Kawara}, K., \& {Okuda}, H. 1998, \nat, 394, 248

\bibitem[{{Barnes} \& {Hernquist}(1996)}]{bar96}
{Barnes}, J.~E. \& {Hernquist}, L. 1996, \apj, 471, 115

\bibitem[{{Barnes} \& {Hernquist}(1991)}]{bar91}
{Barnes}, J.~E. \& {Hernquist}, L.~E. 1991, \apjl, 370, L65

\bibitem[{{Bernes}(1979)}]{ber79}
{Bernes}, C. 1979, \aap, 73, 67

\bibitem[{{Blain} {et~al.}(2004){Blain}, {Chapman}, {Smail}, \&
  {Ivison}}]{bla04}
{Blain}, A.~W., {Chapman}, S.~C., {Smail}, I., \& {Ivison}, R. 2004, \apj, 611,
  725

\bibitem[{{Blain} {et~al.}(1999){Blain}, {Kneib}, {Ivison}, \& {Smail}}]{bla99}
{Blain}, A.~W., {Kneib}, J.-P., {Ivison}, R.~J., \& {Smail}, I. 1999, \apjl,
  512, L87

\bibitem[{{Blain} {et~al.}(2002)}]{bla02}
{Blain}, A.~W. {et~al.} 2002, \physrep, 369, 111

\bibitem[{{Blitz} {et~al.}(2007){Blitz}, {Fukui}, {Kawamura}, {Leroy},
  {Mizuno}, \& {Rosolowsky}}]{bli07}
{Blitz}, L., {Fukui}, Y., {Kawamura}, A., {Leroy}, A., {Mizuno}, N., \&
  {Rosolowsky}, E. 2007, in Protostars and Planets V, ed. B.~{Reipurth},
  D.~{Jewitt}, \& K.~{Keil}, 81--96

\bibitem[{{Blitz} \& {Rosolowsky}(2006)}]{bli06}
{Blitz}, L. \& {Rosolowsky}, E. 2006, \apj, 650, 933

\bibitem[{{Bondi}(1952)}]{bon52}
{Bondi}, H. 1952, \mnras, 112, 195

\bibitem[{{Bondi} \& {Hoyle}(1944)}]{bon44}
{Bondi}, H. \& {Hoyle}, F. 1944, \mnras, 104, 273

\bibitem[{{Borys} {et~al.}(2005){Borys}, {Smail}, {Chapman}, {Blain},
  {Alexander}, \& {Ivison}}]{bor05}
{Borys}, C., {Smail}, I., {Chapman}, S.~C., {Blain}, A.~W., {Alexander}, D.~M.,
  \& {Ivison}, R.~J. 2005, \apj, 635, 853

\bibitem[{{Bouch{\'e}} {et~al.}(2007)}]{bou07}
{Bouch{\'e}}, N. {et~al.} 2007, \apj, 671, 303

\bibitem[{{Bradford} {et~al.}(2004){Bradford}, {Ade}, {Aguirre}, {Bock},
  {Dragovan}, {Duband}, {Earle}, {Glenn}, {Matsuhara}, {Naylor}, {Nguyen},
  {Yun}, \& {Zmuidzinas}}]{bra04}
{Bradford}, C.~M., {Ade}, P.~A.~R., {Aguirre}, J.~E., {Bock}, J.~J.,
  {Dragovan}, M., {Duband}, L., {Earle}, L., {Glenn}, J., {Matsuhara}, H.,
  {Naylor}, B.~J., {Nguyen}, H.~T., {Yun}, M., \& {Zmuidzinas}, J., eds. 2004,
  Presented at the Society of Photo-Optical Instrumentation Engineers (SPIE)
  Conference, Vol. 5498, {Z-Spec: a broadband millimeter-wave grating
  spectrometer: design, construction, and first cryogenic measurements}

\bibitem[{{Bullock} {et~al.}(2001){Bullock}, {Kolatt}, {Sigad}, {Somerville},
  {Kravtsov}, {Klypin}, {Primack}, \& {Dekel}}]{bul01}
{Bullock}, J.~S., {Kolatt}, T.~S., {Sigad}, Y., {Somerville}, R.~S.,
  {Kravtsov}, A.~V., {Klypin}, A.~A., {Primack}, J.~R., \& {Dekel}, A. 2001,
  \mnras, 321, 559

\bibitem[{{Carilli} \& {Wang}(2006)}]{car06}
{Carilli}, C.~L. \& {Wang}, R. 2006, \aj, 131, 2763

\bibitem[{{Castor} {et~al.}(1975){Castor}, {McCray}, \& {Weaver}}]{cas75}
{Castor}, J., {McCray}, R., \& {Weaver}, R. 1975, \apjl, 200, L107

\bibitem[{{Chapman} {et~al.}(2003{\natexlab{a}}){Chapman}, {Blain}, {Ivison},
  \& {Smail}}]{cha03a}
{Chapman}, S.~C., {Blain}, A.~W., {Ivison}, R.~J., \& {Smail}, I.~R.
  2003{\natexlab{a}}, \nat, 422, 695

\bibitem[{{Chapman} {et~al.}(2005){Chapman}, {Blain}, {Smail}, \&
  {Ivison}}]{cha05}
{Chapman}, S.~C., {Blain}, A.~W., {Smail}, I., \& {Ivison}, R.~J. 2005, \apj,
  622, 772

\bibitem[{{Chapman} {et~al.}(2003{\natexlab{b}}){Chapman}, {Windhorst},
  {Odewahn}, {Yan}, \& {Conselice}}]{cha03b}
{Chapman}, S.~C., {Windhorst}, R., {Odewahn}, S., {Yan}, H., \& {Conselice}, C.
  2003{\natexlab{b}}, \apj, 599, 92

\bibitem[{{Coppin} {et~al.}(2008{\natexlab{a}}){Coppin}, {Halpern}, {Scott},
  {Borys}, {Dunlop}, {Dunne}, {Ivison}, {Wagg}, {Aretxaga}, {Battistelli},
  {Benson}, {Blain}, {Chapman}, {Clements}, {Dye}, {Farrah}, {Hughes},
  {Jenness}, {van Kampen}, {Lacey}, {Mortier}, {Pope}, {Priddey}, {Serjeant},
  {Smail}, {Stevens}, \& {Vaccari}}]{cop08b}
{Coppin}, K., {Halpern}, M., {Scott}, D., {Borys}, C., {Dunlop}, J., {Dunne},
  L., {Ivison}, R., {Wagg}, J., {Aretxaga}, I., {Battistelli}, E., {Benson},
  A., {Blain}, A., {Chapman}, S., {Clements}, D., {Dye}, S., {Farrah}, D.,
  {Hughes}, D., {Jenness}, T., {van Kampen}, E., {Lacey}, C., {Mortier}, A.,
  {Pope}, A., {Priddey}, R., {Serjeant}, S., {Smail}, I., {Stevens}, J., \&
  {Vaccari}, M. 2008{\natexlab{a}}, \mnras, 384, 1597

\bibitem[{{Coppin} {et~al.}(2006)}]{cop06}
{Coppin}, K. {et~al.} 2006, \mnras, 372, 1621

\bibitem[{{Coppin} {et~al.}(2008{\natexlab{b}}){Coppin}, {Swinbank}, {Neri},
  {Cox}, {Alexander}, {Smail}, {Page}, {Stevens}, {Knudsen}, {Ivison},
  {Beelen}, {Bertoldi}, \& {Omont}}]{cop08}
{Coppin}, K.~E.~K., {Swinbank}, A.~M., {Neri}, R., {Cox}, P., {Alexander},
  D.~M., {Smail}, I., {Page}, M.~J., {Stevens}, J.~A., {Knudsen}, K.~K.,
  {Ivison}, R.~J., {Beelen}, A., {Bertoldi}, F., \& {Omont}, A.
  2008{\natexlab{b}}, \mnras, 389, 45

\bibitem[{{Cox} {et~al.}(2006){Cox}, {Jonsson}, {Primack}, \&
  {Somerville}}]{cox06a}
{Cox}, T.~J., {Jonsson}, P., {Primack}, J.~R., \& {Somerville}, R.~S. 2006,
  \mnras, 373, 1013

\bibitem[{{Dav{\'e}}(2008)}]{dav08}
{Dav{\'e}}, R. 2008, \mnras, 385, 147

\bibitem[{{Dav{\'e}} {et~al.}(1999){Dav{\'e}}, {Hernquist}, {Katz}, \&
  {Weinberg}}]{dav99}
{Dav{\'e}}, R., {Hernquist}, L., {Katz}, N., \& {Weinberg}, D.~H. 1999, \apj,
  511, 521

\bibitem[{{Di Matteo} {et~al.}(2005){Di Matteo}, {Springel}, \&
  {Hernquist}}]{dim05}
{Di Matteo}, T., {Springel}, V., \& {Hernquist}, L. 2005, \nat, 433, 604

\bibitem[{{Downes} \& {Solomon}(1998)}]{dow98}
{Downes}, D. \& {Solomon}, P.~M. 1998, \apj, 507, 615

\bibitem[{{Downes} \& {Solomon}(2003)}]{dow03}
---. 2003, \apj, 582, 37

\bibitem[{{Draine} \& {Li}(2007)}]{dra07}
{Draine}, B.~T. \& {Li}, A. 2007, \apj, 657, 810

\bibitem[{{Dwek}(1998)}]{dwe98}
{Dwek}, E. 1998, \apj, 501, 643

\bibitem[{{Frayer} {et~al.}(1999){Frayer}, {Ivison}, {Scoville}, {Evans},
  {Yun}, {Smail}, {Barger}, {Blain}, \& {Kneib}}]{fra99}
{Frayer}, D.~T., {Ivison}, R.~J., {Scoville}, N.~Z., {Evans}, A.~S., {Yun},
  M.~S., {Smail}, I., {Barger}, A.~J., {Blain}, A.~W., \& {Kneib}, J.-P. 1999,
  \apjl, 514, L13

\bibitem[{{Frayer} {et~al.}(1998){Frayer}, {Ivison}, {Scoville}, {Yun},
  {Evans}, {Smail}, {Blain}, \& {Kneib}}]{fra98}
{Frayer}, D.~T., {Ivison}, R.~J., {Scoville}, N.~Z., {Yun}, M., {Evans}, A.~S.,
  {Smail}, I., {Blain}, A.~W., \& {Kneib}, J.-P. 1998, \apjl, 506, L7

\bibitem[{{Genzel} {et~al.}(2003){Genzel}, {Baker}, {Tacconi}, {Lutz}, {Cox},
  {Guilloteau}, \& {Omont}}]{gen03}
{Genzel}, R., {Baker}, A.~J., {Tacconi}, L.~J., {Lutz}, D., {Cox}, P.,
  {Guilloteau}, S., \& {Omont}, A. 2003, \apj, 584, 633

\bibitem[{{Greve} {et~al.}(2003){Greve}, {Ivison}, \& {Papadopoulos}}]{gre03}
{Greve}, T.~R., {Ivison}, R.~J., \& {Papadopoulos}, P.~P. 2003, \apj, 599, 839

\bibitem[{{Greve} {et~al.}(2005)}]{gre05}
{Greve}, T.~R. {et~al.} 2005, \mnras, 359, 1165

\bibitem[{{Groves} {et~al.}(2008){Groves}, {Dopita}, {Sutherland}, {Kewley},
  {Fischera}, {Leitherer}, {Brandl}, \& {van Breugel}}]{gro08}
{Groves}, B., {Dopita}, M.~A., {Sutherland}, R.~S., {Kewley}, L.~J.,
  {Fischera}, J., {Leitherer}, C., {Brandl}, B., \& {van Breugel}, W. 2008,
  \apjs, 176, 438

\bibitem[{{Groves} {et~al.}(2004){Groves}, {Dopita}, \& {Sutherland}}]{gro04}
{Groves}, B.~A., {Dopita}, M.~A., \& {Sutherland}, R.~S. 2004, \apjs, 153, 9

\bibitem[{{Hailey-Dunsheath} {et~al.}(2008){Hailey-Dunsheath}, {Nikola},
  {Oberst}, {Parshley}, {Stacey}, {Farrah}, {Benford}, \& {Staguhn}}]{hai08}
{Hailey-Dunsheath}, S., {Nikola}, T., {Oberst}, T., {Parshley}, S., {Stacey},
  G.~J., {Farrah}, D., {Benford}, D.~J., \& {Staguhn}, J. 2008, in EAS
  Publications Series, Vol.~31, EAS Publications Series, ed. C.~{Kramer},
  S.~{Aalto}, \& R.~{Simon}, 159--162

\bibitem[{{Hainline} {et~al.}(2006){Hainline}, {Blain}, {Greve}, {Chapman},
  {Smail}, \& {Ivison}}]{hai06}
{Hainline}, L.~J., {Blain}, A.~W., {Greve}, T.~R., {Chapman}, S.~C., {Smail},
  I., \& {Ivison}, R.~J. 2006, \apj, 650, 614

\bibitem[{{Harris} {et~al.}(2007){Harris}, {Baker}, {Jewell}, {Rauch}, {Zonak},
  {O'Neil}, {Shelton}, {Norrod}, {Ray}, \& {Watts}}]{har07}
{Harris}, A.~I., {Baker}, A.~J., {Jewell}, P.~R., {Rauch}, K.~P., {Zonak},
  S.~G., {O'Neil}, K., {Shelton}, A.~L., {Norrod}, R.~D., {Ray}, J., \&
  {Watts}, G. 2007, in Astronomical Society of the Pacific Conference Series,
  Vol. 375, From Z-Machines to ALMA: (Sub)Millimeter Spectroscopy of Galaxies,
  ed. A.~J. {Baker}, J.~{Glenn}, A.~I. {Harris}, J.~G. {Mangum}, \& M.~S.
  {Yun}, 82--+

\bibitem[{{Hernquist}(1990)}]{her90}
{Hernquist}, L. 1990, \apj, 356, 359

\bibitem[{{Ho}(2007)}]{ho07}
{Ho}, L.~C. 2007, \apj, 669, 821

\bibitem[{{Hopkins} {et~al.}(2009){Hopkins}, {Bundy}, {Croton}, {Hernquist},
  {Keres}, {Khochfar}, {Stewart}, {Wetzel}, \& {Younger}}]{hop09d}
{Hopkins}, P.~F., {Bundy}, K., {Croton}, D., {Hernquist}, L., {Keres}, D.,
  {Khochfar}, S., {Stewart}, K., {Wetzel}, A., \& {Younger}, J.~D. 2009, ArXiv
  e-prints

\bibitem[{{Hopkins} {et~al.}(2008){Hopkins}, {Hernquist}, {Cox}, {Younger}, \&
  {Besla}}]{hop08d}
{Hopkins}, P.~F., {Hernquist}, L., {Cox}, T.~J., {Younger}, J.~D., \& {Besla},
  G. 2008, \apj, 688, 757

\bibitem[{{Hopkins} {et~al.}(2007){Hopkins}, {Richards}, \&
  {Hernquist}}]{hop07}
{Hopkins}, P.~F., {Richards}, G.~T., \& {Hernquist}, L. 2007, \apj, 654, 731

\bibitem[{{Hopkins} {et~al.}(2005)}]{hop05a}
{Hopkins}, P.~F. {et~al.} 2005, \apj, 630, 705

\bibitem[{{Hopkins} {et~al.}(2006)}]{hop06}
---. 2006, \apjs, 163, 1

\bibitem[{{Hughes} {et~al.}(1998)}]{hug98}
{Hughes}, D.~H. {et~al.} 1998, \nat, 394, 241

\bibitem[{{Iono} {et~al.}(2009){Iono}, {Wilson}, {Yun}, {Baker}, {Petitpas},
  {Peck}, {Krips}, {Cox}, {Matsushita}, {Mihos}, \& {Pihlstrom}}]{ion09}
{Iono}, D., {Wilson}, C.~D., {Yun}, M.~S., {Baker}, A.~J., {Petitpas}, G.~R.,
  {Peck}, A.~B., {Krips}, M., {Cox}, T.~J., {Matsushita}, S., {Mihos}, J.~C.,
  \& {Pihlstrom}, Y. 2009, \apj, 695, 1537

\bibitem[{{Ivison} {et~al.}(2002){Ivison}, {Greve}, {Smail}, {Dunlop}, {Roche},
  {Scott}, {Page}, {Stevens}, {Almaini}, {Blain}, {Willott}, {Fox}, {Gilbank},
  {Serjeant}, \& {Hughes}}]{ivi02}
{Ivison}, R.~J., {Greve}, T.~R., {Smail}, I., {Dunlop}, J.~S., {Roche}, N.~D.,
  {Scott}, S.~E., {Page}, M.~J., {Stevens}, J.~A., {Almaini}, O., {Blain},
  A.~W., {Willott}, C.~J., {Fox}, M.~J., {Gilbank}, D.~G., {Serjeant}, S., \&
  {Hughes}, D.~H. 2002, \mnras, 337, 1

\bibitem[{{Ivison} {et~al.}(2001){Ivison}, {Smail}, {Frayer}, {Kneib}, \&
  {Blain}}]{ivi01}
{Ivison}, R.~J., {Smail}, I., {Frayer}, D.~T., {Kneib}, J.-P., \& {Blain},
  A.~W. 2001, \apjl, 561, L45

\bibitem[{{Jonsson}(2006)}]{jon06a}
{Jonsson}, P. 2006, \mnras, 372, 2

\bibitem[{{Jonsson} {et~al.}(2006){Jonsson}, {Cox}, {Primack}, \&
  {Somerville}}]{jon06b}
{Jonsson}, P., {Cox}, T.~J., {Primack}, J.~R., \& {Somerville}, R.~S. 2006,
  \apj, 637, 255

\bibitem[{{Jonsson} {et~al.}(2009){Jonsson}, {Groves}, \& {Cox}}]{jon09}
{Jonsson}, P., {Groves}, B., \& {Cox}, T.~J. 2009, ArXiv e-prints

\bibitem[{{Juvela}(2005)}]{juv05}
{Juvela}, M. 2005, \aap, 440, 531

\bibitem[{{Katz} {et~al.}(1996){Katz}, {Weinberg}, \& {Hernquist}}]{kat96}
{Katz}, N., {Weinberg}, D.~H., \& {Hernquist}, L. 1996, \apjs, 105, 19

\bibitem[{{Kennicutt}(1998{\natexlab{a}})}]{ken98a}
{Kennicutt}, Jr., R.~C. 1998{\natexlab{a}}, \araa, 36, 189

\bibitem[{{Kennicutt}(1998{\natexlab{b}})}]{ken98b}
---. 1998{\natexlab{b}}, \apj, 498, 541

\bibitem[{{Kennicutt} {et~al.}(2007){Kennicutt}, {Calzetti}, {Walter}, {Helou},
  {Hollenbach}, {Armus}, {Bendo}, {Dale}, {Draine}, {Engelbracht}, {Gordon},
  {Prescott}, {Regan}, {Thornley}, {Bot}, {Brinks}, {de Blok}, {de Mello},
  {Meyer}, {Moustakas}, {Murphy}, {Sheth}, \& {Smith}}]{ken07}
{Kennicutt}, Jr., R.~C., {Calzetti}, D., {Walter}, F., {Helou}, G.,
  {Hollenbach}, D.~J., {Armus}, L., {Bendo}, G., {Dale}, D.~A., {Draine},
  B.~T., {Engelbracht}, C.~W., {Gordon}, K.~D., {Prescott}, M.~K.~M., {Regan},
  M.~W., {Thornley}, M.~D., {Bot}, C., {Brinks}, E., {de Blok}, E., {de Mello},
  D., {Meyer}, M., {Moustakas}, J., {Murphy}, E.~J., {Sheth}, K., \& {Smith},
  J.~D.~T. 2007, \apj, 671, 333

\bibitem[{{Keres} {et~al.}(2003){Keres}, {Yun}, \& {Young}}]{ker03}
{Keres}, D., {Yun}, M.~S., \& {Young}, J.~S. 2003, \apj, 582, 659

\bibitem[{{Kov{\'a}cs} {et~al.}(2006)}]{kov06}
{Kov{\'a}cs}, A. {et~al.} 2006, \apj, 650, 592

\bibitem[{{Kroupa}(2002)}]{kro02}
{Kroupa}, P. 2002, Science, 295, 82

\bibitem[{{Leitherer} {et~al.}(1999)}]{lei99}
{Leitherer}, C. {et~al.} 1999, \apjs, 123, 3

\bibitem[{{Li} {et~al.}(2007){Li}, {Hernquist}, {Robertson}, {Cox}, {Hopkins},
  {Springel}, {Gao}, {Di Matteo}, {Zentner}, {Jenkins}, \& {Yoshida}}]{li07}
{Li}, Y., {Hernquist}, L., {Robertson}, B., {Cox}, T.~J., {Hopkins}, P.~F.,
  {Springel}, V., {Gao}, L., {Di Matteo}, T., {Zentner}, A.~R., {Jenkins}, A.,
  \& {Yoshida}, N. 2007, \apj, 665, 187

\bibitem[{{Lidz} {et~al.}(2006){Lidz}, {Hopkins}, {Cox}, {Hernquist}, \&
  {Robertson}}]{lid06}
{Lidz}, A., {Hopkins}, P.~F., {Cox}, T.~J., {Hernquist}, L., \& {Robertson}, B.
  2006, \apj, 641, 41

\bibitem[{{McKee} \& {Ostriker}(1977)}]{mck77}
{McKee}, C.~F. \& {Ostriker}, J.~P. 1977, \apj, 218, 148

\bibitem[{{Men{\'e}ndez-Delmestre} {et~al.}(2007){Men{\'e}ndez-Delmestre},
  {Blain}, {Alexander}, {Smail}, {Armus}, {Chapman}, {Frayer}, {Ivison}, \&
  {Teplitz}}]{men07}
{Men{\'e}ndez-Delmestre}, K., {Blain}, A.~W., {Alexander}, D.~M., {Smail}, I.,
  {Armus}, L., {Chapman}, S.~C., {Frayer}, D.~T., {Ivison}, R.~J., \&
  {Teplitz}, H.~I. 2007, \apjl, 655, L65

\bibitem[{{Mihos} \& {Hernquist}(1994)}]{mih94a}
{Mihos}, J.~C. \& {Hernquist}, L. 1994, \apjl, 431, L9

\bibitem[{{Mihos} \& {Hernquist}(1996)}]{mih96}
---. 1996, \apj, 464, 641

\bibitem[{{Narayanan} {et~al.}(2008{\natexlab{a}}){Narayanan}, {Cox}, {Kelly},
  {Dav{\'e}}, {Hernquist}, {Di Matteo}, {Hopkins}, {Kulesa}, {Robertson}, \&
  {Walker}}]{nar08a}
{Narayanan}, D., {Cox}, T.~J., {Kelly}, B., {Dav{\'e}}, R., {Hernquist}, L.,
  {Di Matteo}, T., {Hopkins}, P.~F., {Kulesa}, C., {Robertson}, B., \&
  {Walker}, C.~K. 2008{\natexlab{a}}, \apjs, 176, 331

\bibitem[{{Narayanan} {et~al.}(2006{\natexlab{a}}){Narayanan}, {Cox},
  {Robertson}, {Dav{\'e}}, {Di Matteo}, {Hernquist}, {Hopkins}, {Kulesa}, \&
  {Walker}}]{nar06a}
{Narayanan}, D., {Cox}, T.~J., {Robertson}, B., {Dav{\'e}}, R., {Di Matteo},
  T., {Hernquist}, L., {Hopkins}, P., {Kulesa}, C., \& {Walker}, C.~K.
  2006{\natexlab{a}}, \apjl, 642, L107

\bibitem[{{Narayanan} {et~al.}(2008{\natexlab{b}}){Narayanan}, {Cox},
  {Shirley}, {Dav{\'e}}, {Hernquist}, \& {Walker}}]{nar08b}
{Narayanan}, D., {Cox}, T.~J., {Shirley}, Y., {Dav{\'e}}, R., {Hernquist}, L.,
  \& {Walker}, C.~K. 2008{\natexlab{b}}, \apj, 684, 996

\bibitem[{{Narayanan} {et~al.}(2009){Narayanan}, {Hayward}, {Cox}, {Hernquist},
  {Jonsson}, {Younger}, \& {Groves}}]{nar09}
{Narayanan}, D., {Hayward}, C.~C., {Cox}, T.~J., {Hernquist}, L., {Jonsson},
  P., {Younger}, J.~D., \& {Groves}, B. 2009, ArXiv e-prints: arXiv/0904.0004

\bibitem[{{Narayanan} {et~al.}(2006{\natexlab{b}}){Narayanan}, {Kulesa},
  {Boss}, \& {Walker}}]{nar06b}
{Narayanan}, D., {Kulesa}, C.~A., {Boss}, A., \& {Walker}, C.~K.
  2006{\natexlab{b}}, \apj, 647, 1426

\bibitem[{{Narayanan} {et~al.}(2008{\natexlab{c}}){Narayanan}, {Li}, {Cox},
  {Hernquist}, {Hopkins}, {Chakrabarti}, {Dav{\'e}}, {Di Matteo}, {Gao},
  {Kulesa}, {Robertson}, \& {Walker}}]{nar08c}
{Narayanan}, D., {Li}, Y., {Cox}, T.~J., {Hernquist}, L., {Hopkins}, P.,
  {Chakrabarti}, S., {Dav{\'e}}, R., {Di Matteo}, T., {Gao}, L., {Kulesa}, C.,
  {Robertson}, B., \& {Walker}, C.~K. 2008{\natexlab{c}}, \apjs, 174, 13

\bibitem[{{Neri} {et~al.}(2003){Neri}, {Genel}, {Ivison}, {Bertoldi}, {Blain},
  {Chapman}, {Cox}, {Greve}, {Omont}, \& {Frayer}}]{ner03}
{Neri}, R., {Genel}, R., {Ivison}, R.~J., {Bertoldi}, F., {Blain}, A.~W.,
  {Chapman}, S.~C., {Cox}, P., {Greve}, T.~R., {Omont}, A., \& {Frayer}, D.~T.
  2003, \apjl, 597, L113

\bibitem[{{Papadopoulos} \& {Ivison}(2002)}]{pap02}
{Papadopoulos}, P.~P. \& {Ivison}, R.~J. 2002, \apjl, 564, L9

\bibitem[{{Pascucci} {et~al.}(2004){Pascucci}, {Wolf}, {Steinacker},
  {Dullemond}, {Henning}, {Niccolini}, {Woitke}, \& {Lopez}}]{pas04}
{Pascucci}, I., {Wolf}, S., {Steinacker}, J., {Dullemond}, C.~P., {Henning},
  T., {Niccolini}, G., {Woitke}, P., \& {Lopez}, B. 2004, \aap, 417, 793

\bibitem[{{Pelupessy} {et~al.}(2006){Pelupessy}, {Papadopoulos}, \& {van der
  Werf}}]{pel06}
{Pelupessy}, F.~I., {Papadopoulos}, P.~P., \& {van der Werf}, P. 2006, \apj,
  645, 1024

\bibitem[{{Robertson} {et~al.}(2006){Robertson}, {Hernquist}, {Cox}, {Di
  Matteo}, {Hopkins}, {Martini}, \& {Springel}}]{rob06b}
{Robertson}, B., {Hernquist}, L., {Cox}, T.~J., {Di Matteo}, T., {Hopkins},
  P.~F., {Martini}, P., \& {Springel}, V. 2006, \apj, 641, 90

\bibitem[{{Robertson} \& {Kravtsov}(2007)}]{rob07b}
{Robertson}, B. \& {Kravtsov}, A. 2007, ArXiv e-prints, 710

\bibitem[{{Robertson} {et~al.}(2004){Robertson}, {Yoshida}, {Springel}, \&
  {Hernquist}}]{rob04}
{Robertson}, B., {Yoshida}, N., {Springel}, V., \& {Hernquist}, L. 2004, \apj,
  606, 32

\bibitem[{{Sakamoto} {et~al.}(1999)}]{sak99}
{Sakamoto}, K. {et~al.} 1999, \apj, 514, 68

\bibitem[{{Schmidt}(1959)}]{sch59}
{Schmidt}, M. 1959, \apj, 129, 243

\bibitem[{{Sch{\"o}ier} {et~al.}(2005){Sch{\"o}ier}, {van der Tak}, {van
  Dishoeck}, \& {Black}}]{sch05}
{Sch{\"o}ier}, F.~L., {van der Tak}, F.~F.~S., {van Dishoeck}, E.~F., \&
  {Black}, J.~H. 2005, \aap, 432, 369

\bibitem[{{Shapley} {et~al.}(2004){Shapley}, {Erb}, {Pettini}, {Steidel}, \&
  {Adelberger}}]{sha04}
{Shapley}, A.~E., {Erb}, D.~K., {Pettini}, M., {Steidel}, C.~C., \&
  {Adelberger}, K.~L. 2004, \apj, 612, 108

\bibitem[{{Solomon} \& {Barrett}(1991)}]{sol91}
{Solomon}, P.~M. \& {Barrett}, J.~W. 1991, in IAU Symposium, Vol. 146, Dynamics
  of Galaxies and Their Molecular Cloud Distributions, ed. F.~{Combes} \&
  F.~{Casoli}, 235--+

\bibitem[{{Solomon} \& {Vanden Bout}(2005)}]{sol05}
{Solomon}, P.~M. \& {Vanden Bout}, P.~A. 2005, \araa, 43, 677

\bibitem[{{Springel}(2005)}]{spr05b}
{Springel}, V. 2005, \mnras, 364, 1105

\bibitem[{{Springel} {et~al.}(2005){Springel}, {Di Matteo}, \&
  {Hernquist}}]{spr05a}
{Springel}, V., {Di Matteo}, T., \& {Hernquist}, L. 2005, \mnras, 361, 776

\bibitem[{{Springel} \& {Hernquist}(2002)}]{spr02}
{Springel}, V. \& {Hernquist}, L. 2002, \mnras, 333, 649

\bibitem[{{Springel} \& {Hernquist}(2003)}]{spr03a}
---. 2003, \mnras, 339, 289

\bibitem[{{Swinbank} {et~al.}(2004)}]{swi04}
{Swinbank}, A.~M. {et~al.} 2004, \apj, 617, 64

\bibitem[{{Swinbank} {et~al.}(2008)}]{swi08}
---. 2008, \mnras, 391, 420

\bibitem[{{Tacconi} {et~al.}(2006)}]{tac06}
{Tacconi}, L.~J. {et~al.} 2006, \apj, 640, 228

\bibitem[{{Tacconi} {et~al.}(2008)}]{tac08}
---. 2008, \apj, 680, 246

\bibitem[{{Valiante} {et~al.}(2007){Valiante}, {Lutz}, {Sturm}, {Genzel},
  {Tacconi}, {Lehnert}, \& {Baker}}]{val07}
{Valiante}, E., {Lutz}, D., {Sturm}, E., {Genzel}, R., {Tacconi}, L.~J.,
  {Lehnert}, M.~D., \& {Baker}, A.~J. 2007, \apj, 660, 1060

\bibitem[{{van Dokkum}(2008)}]{van08}
{van Dokkum}, P.~G. 2008, \apj, 674, 29

\bibitem[{{V{\'a}zquez} \& {Leitherer}(2005)}]{vaz05}
{V{\'a}zquez}, G.~A. \& {Leitherer}, C. 2005, \apj, 621, 695

\bibitem[{{Weingartner} \& {Draine}(2001)}]{wei01}
{Weingartner}, J.~C. \& {Draine}, B.~T. 2001, \apj, 548, 296

\bibitem[{{Wei{\ss}} {et~al.}(2007){Wei{\ss}}, {Downes}, {Walter}, \&
  {Henkel}}]{wei07}
{Wei{\ss}}, A., {Downes}, D., {Walter}, F., \& {Henkel}, C. 2007, in
  Astronomical Society of the Pacific Conference Series, Vol. 375, From
  Z-Machines to ALMA: (Sub)Millimeter Spectroscopy of Galaxies, ed. A.~J.
  {Baker}, J.~{Glenn}, A.~I. {Harris}, J.~G. {Mangum}, \& M.~S. {Yun}, 25--+

\bibitem[{{Wei{\ss}} {et~al.}(2005){Wei{\ss}}, {Walter}, \& {Scoville}}]{wei05}
{Wei{\ss}}, A., {Walter}, F., \& {Scoville}, N.~Z. 2005, \aap, 438, 533

\bibitem[{{Wilson} {et~al.}(2008){Wilson}, {Austermann}, {Perera}, {Scott},
  {Ade}, {Bock}, {Glenn}, {Golwala}, {Kim}, {Kang}, {Lydon}, {Mauskopf},
  {Predmore}, {Roberts}, {Souccar}, \& {Yun}}]{wil08a}
{Wilson}, G.~W., {Austermann}, J.~E., {Perera}, T.~A., {Scott}, K.~S., {Ade},
  P.~A.~R., {Bock}, J.~J., {Glenn}, J., {Golwala}, S.~R., {Kim}, S., {Kang},
  Y., {Lydon}, D., {Mauskopf}, P.~D., {Predmore}, C.~R., {Roberts}, C.~M.,
  {Souccar}, K., \& {Yun}, M.~S. 2008, \mnras, 386, 807

\bibitem[{{Younger} {et~al.}(2008)}]{you08b}
{Younger}, J.~D. {et~al.} 2008, \apj, 688, 59

\end{thebibliography}
\end{document}